\documentclass[acmlarge,nonacm]{acmart} 

\usepackage{natbib}
\usepackage{algorithm}
\usepackage{algpseudocode}

\AtBeginDocument{%
  }

\begin{document}

\title{Tactical Asset Allocation with Macroeconomic Regime Detection}

\author{Daniel Cunha Oliveira}
\affiliation{%
  \institution{University of São Paulo - Institute of Mathematics and Statistics}
  \city{São Paulo}
  \country{Brazil}
}

\author{Dylan Sandfelder}
\authornote{Corresponding author.}
\email{dylan.sandfelder@eng.ox.ac.uk}
\affiliation{%
  \institution{University of Oxford - Oxford-Man Institute of Quantitative Finance}
  \city{Oxford}
  \country{UK}
}

\author{André Fujita}
\affiliation{%
  \institution{University of São Paulo - Institute of Mathematics and Statistics}
  \city{São Paulo}
  \country{Brazil}
}
\affiliation{%
  \institution{Kyushu University - Medical Institute of Bioregulation}
  \city{Fukuoka}
  \country{Japan}
}

\author{Xiaowen Dong}
\affiliation{%
  \institution{University of Oxford - Oxford-Man Institute of Quantitative Finance}
  \city{Oxford}
  \country{UK}
}

\author{Mihai Cucuringu}
\affiliation{%
  \institution{University of California Los Angeles - Department of Mathematics}
  \city{Los Angeles}
  \state{California}
  \country{USA}
}
\affiliation{%
  \institution{University of Oxford - Oxford-Man Institute of Quantitative Finance}
  \city{Oxford}
  \country{UK}
}
\affiliation{%
  \institution{University of Oxford - Department of Statistics}
  \city{Oxford}
  \country{UK}
}

\renewcommand{\shortauthors}{Oliveira et al.}

\begin{abstract}
This paper extends the tactical asset allocation literature by incorporating regime modeling using techniques from machine learning. We propose a novel model that classifies current regimes, forecasts the distribution of future regimes, and integrates these forecasts with the historical performance of individual assets to optimize portfolio allocations. Utilizing a macroeconomic data set from the FRED-MD database, our approach employs a modified k-means algorithm to ensure consistent regime classification over time. We then leverage these regime predictions to estimate expected returns and volatilities, which are subsequently mapped into portfolio allocations using various sizing schemes. Our method outperforms traditional benchmarks such as equal-weight, buy-and-hold, and random regime models. Additionally, we are the first to apply a regime detection model from a large macroeconomic dataset to tactical asset allocation, demonstrating significant improvements in portfolio performance. Our work presents several key contributions, including a novel data-driven regime detection algorithm tailored for uncertainty in forecasted regimes and applying the FRED-MD data set for tactical asset allocation.
\end{abstract}



\received{14 March 2025}

\maketitle

\section{Introduction}

The asset allocation literature primarily concerns building optimal portfolios tailored to the utility functions of the investors. The most popular framework for asset allocation and portfolio choice is the seminal work of \cite{markowitz-1953}. Known as mean-variance optimization (MVO), this method estimates expected returns and covariances for financial assets. Then, it uses these estimates to solve an optimization problem to determine portfolio weights. This method is widely used in practice to maximize returns for a given level of portfolio risk.

The tactical asset allocation literature follows Markowitz's work closely, but focuses on signals that can be used as tactical portfolio tilts to avoid tail risk scenarios. A typical tactical asset allocation model consists of three steps: (1) building signals to model the state of the economy and/or assets, (2) using the time series of asset prices and/or the signals constructed in the previous step to develop individual asset forecasts, and (3) mapping these forecasts into portfolio weights or allocations. Among the various signals used for tactical asset allocation, outputs from models that account for regime dynamics are particularly noteworthy, as shown in \cite{guidolin-timmermann-2007} and \cite{ang-bekaert-2004}. These models aim to estimate probabilities for a given number of unobserved states (regimes) using observed data.

The regime modeling literature is extensive and has been explored from both econometric and machine learning perspectives. From an econometric point of view, the seminal work of \cite{hamilton-1989} laid the foundation for subsequent studies. This strand of literature has primarily focused on regime-switching models, typically with a specified number of estimated states. In contrast, the machine learning literature mostly relies on clustering techniques, as in \cite{chen2021clustering} and \cite{prakash2021structural}. A recent study by \cite{chen-etal-2023} combines Generative Adversarial Networks (GANs) and Recurrent Neural Networks (RNNs) to underscore the importance of modeling macroeconomic states for predicting individual stock returns. The authors provide evidence that four distinct macroeconomic states effectively capture the dynamics of stock returns.

Both strands have been successfully applied to various tasks, including stock return forecasting, volatility forecasting, asset allocation, and tactical asset allocation. However, because both approaches use time series of asset returns as the main inputs for their models, they are forced to deal with noisy data. This can negatively impact the quality of the regimes uncovered in the data by introducing excessive uncertainty, thus rendering  the models less accurate.

This paper extends the tactical asset allocation literature by incorporating macroeconomic regime modeling into the asset allocation process in the form of weighted priors, which effectively addresses some of the issues caused by the noisiness of asset returns data. We extract regime classifications from more stable macroeconomic data rather than asset returns using data-driven clustering techniques from the machine learning literature. We propose a novel model that classifies current regimes, forecasts the distribution of regimes for the next time step, integrates these forecasts with the historical performance of individual assets, and translates the final predictions into portfolio allocations.

Our work uses tools from the unsupervised machine learning literature, particularly clustering methods, to derive the regime classification distributions that we use as effective priors for asset allocation. These tools can be broadly divided into algorithmic and model-based approaches. Algorithmic approaches, such as hierarchical, partition-based, and density-based clustering, use iterative processes to classify entities based on similarity measures. A well-known partition-based method is the k-means algorithm \citep{hartigan-wong-1979}, while DBSCAN is a popular density-based method, effective for discovering arbitrary shape clusters and handling noise \citep{ester-etal-1996, schubert-etal-2017}. 

Our approach is particularly related to fuzzy clustering methods, which extend traditional clustering by allowing data points to belong to multiple clusters with varying degrees of membership. This concept, first introduced by \cite{dunn-1973} and later expanded by \cite{bezdek-1981}, provides a more nuanced view of cluster assignments compared to hard clustering methods and allows us to compute classification probability distributions, which is needed for quantifying uncertainty, a crucial aspect of integrating regime modeling into asset allocation. In fuzzy clustering, rather than assigning each observation to exactly one cluster, the algorithm assigns membership probabilities across all clusters. This connects naturally with model-based clustering approaches, which rely on probabilistic models, typically using mixture models like Gaussian mixture models (GMMs), to classify data by assuming that each cluster follows a distinct probability distribution \citep{fraley-raftery-2002}.

Our asset allocation model can be broken down into three main stages. The first stage of our model employs a novel macroeconomic dataset from the FRED-MD database, consisting of more than 100 monthly time series that characterize the state of the US economy. These time series are fed into a modified version of the k-means algorithm. Similarly to the fuzzy c-means model \cite{bezdek-1981}, our algorithm computes probabilities from any clustering technique based on centroid distances. We leverage this method as a wrapper around the k-means algorithm, defining the modified k-means. The modified k-means algorithm then consists of a two-step approach to classify typical and atypical months, and a matching clustering algorithm to ensure consistency of cluster interpretation over time. The output of this stage is a prediction for the current regime and a probability distribution over regimes for the next time step.

In the second stage, we use these regime predictions to build estimates of expected returns and/or volatilities. This can be achieved using the conditional expected values given the most likely regimes for the next time step based on past regime occurrences, or a simple ridge regression model estimated with data from the same regime. This process results in forecasts of individual asset returns and/or volatilities.

Finally, the last stage involves mapping these forecasts into asset allocations. We experiment with different configurations of sizing schemes, including equal-weight, long-only, long-and-short, and long with a tactical short tilt when a high conviction forecast of an impending economic crisis is present.

By pairing this regime classification model with existing portfolio optimization models, we are able to drive much higher and statistically significant returns even when using relatively simple portfolio optimization models. This empirically demonstrates the value of modeling macroeconomic regime shifts and incorporating that model into a larger investment framework.

\textbf{Main Contributions.} Our work makes several key contributions to the literature on tactical asset allocation and regime modeling. First, we evaluate different methods for computing regime (cluster) uncertainty and demonstrate that a modified k-means algorithm, similar to fuzzy c-means approaches, is better suited for applications requiring smooth, nuanced regime probability transitions over time and overcomes the constraints of deterministic methods. Second, we introduce the use of the comprehensive FRED-MD macroeconomic dataset in a data-driven regime detection model for tactical asset allocation, moving beyond market data alone used by most existing methods and incorporating broader external economic signals. Third, we demonstrate that regime-based portfolios significantly outperform those constructed using random regime classifications. Finally, we show that our method generates valuable signals that lead to superior performance compared to equal-weight, buy-and-hold, and mean-variance optimization strategies.

\section{Related Work}

Our work relates to and contributes to several strands of literature, particularly regime modeling, both from an econometrics and machine learning perspective and asset allocation.

Practitioners have extensively used the concept of regime modeling to capture the complexities of financial markets. Regimes are intuitive because they reflect the different phases or states of an economic or financial process, such as periods of high and low volatility, or other stages of the business cycle. \cite{ang-timmermann-2011} showed that models that account for regimes can capture critical stylized facts about financial time series, such as fat tails, time-varying correlations, skewness, and nonlinearities.

Regime modeling in econometrics has a long tradition, with the seminal work by \cite{hamilton-1989} laying the foundation for Markov-switching models. These models have been widely used to analyze many economic and financial processes. Regime-switching models can be applied to stock market prediction modeling, as demonstrated in the works of \cite{schwert-1989}, \cite{hassan-nath-2005}, \cite{gupta-dhingra-2012}, and \cite{nguyen-2018}. Additionally, regime-switching models have been applied to volatility forecasting, with significant contributions from \cite{hamilton-susmel-1994} and \cite{haas-etal-2004}. Moreover, these models have been used in monetary policy analysis, as explored by \cite{sims-zha-2006}.

In machine learning, the literature has similarly focused on using asset returns as inputs for regime models. Clustering techniques, such as k-means, have been employed to identify regimes within asset prices for various purposes. For instance, \cite{golosnoy-etal-2014}, \cite{andrada_felix-etal-2016}, and \cite{cartea-etal-2023} have used various forms of similarity learning or clustering for volatility forecasting. \cite{miori-cucuringu-2022,lead_lag_ML_Bennett_2022} have applied clustering to lead-lag analysis, while \cite{shu-etal-2024} have explored its application for asset allocation using the mean-variance optimization framework.

The asset allocation literature primarily concerns optimizing portfolio returns while managing risk. Regime shifts significantly impact asset allocation decisions by altering the risk-return profiles of assets. \cite{ang-bekaert-2004} demonstrated how regime changes can influence asset allocation strategies. \cite{guidolin-timmermann-2007} provided a comprehensive asset allocation analysis under multivariate regime-switching models. \cite{tu2010regime} argued that ignoring regime switches can be costly, potentially reducing portfolio returns by 2\% per year. The tactical asset allocation literature, closer to what we propose here, focuses on tactical tilts in portfolios to enhance their robustness against adverse periods and minimize drawdowns. This differs from the classical asset allocation literature, which emphasizes long-term strategic asset allocation. \cite{kritzman-etal-2012} discussed the implications of regime shifts for tactical asset allocation by modeling regimes with a Markov-switching model.

Most existing methods for incorporating regime shifts into asset allocation rely on detecting distributional shifts in the returns data. Although effective for detecting change points, these approaches are highly sensitive to the choice of market data and fail to systematically define regimes based on fundamental market factors. In this work, we bridge the gap between macroeconomic and market data to construct a regime model that integrates external economic signals. This design ensures our model’s applicability across asset allocation frameworks while providing regimes that are interpretable in terms of fundamental macroeconomic drivers.

Our work also uses tools from the unsupervised machine learning and clustering methods literature. Clustering methods are commonly used to identify underlying patterns or regimes within data, and the literature on clustering can be broadly divided into algorithmic and model-based approaches.

Algorithmic approaches rely on iterative processes that classify entities based on similarity measures and are deterministic. These methods do not provide probabilities for regime membership; instead, they assign entities to specific clusters. Common techniques within this category include hierarchical clustering, partition-based algorithms, and density-based algorithms. For example, the k-means algorithm \cite{hartigan-wong-1979} is a widely used partition-based method that minimizes within-cluster variance through an iterative process. At the same time, DBSCAN is a popular density-based method that excels at identifying clusters of arbitrary shapes and handling noise \citep{ester-etal-1996, schubert-etal-2017}.

Fuzzy clustering methods, particularly the fuzzy c-means algorithm introduced by \cite{dunn-1973} and refined by \cite{bezdek-1981}, bridge the gap between deterministic and probabilistic approaches. Unlike traditional clustering methods that assign each observation to exactly one cluster, fuzzy c-means allows observations to belong to multiple clusters with varying degrees of membership. This is achieved through an optimization problem that minimizes the weighted sum of squared distances between points and cluster centers, where the weights represent membership degrees. The resulting membership values provide a natural measure of uncertainty in cluster assignments.

In contrast, model-based clustering relies on probabilistic models that assume each cluster follows a distinct probability distribution, allowing for the estimation of probabilities for each entity's membership across clusters. This probabilistic nature makes model-based approaches, such as GMMs \citep{fraley-raftery-2002}, particularly suitable for regime identification in financial markets, where capturing uncertainty in cluster assignments is valuable.

\section{Methodology}
\label{sec:regimes}

We introduce a data-driven regime detection method that defines monthly regime distributions based on macroeconomic variables. Our approach begins with a regime classification strategy that leverages k-means clustering with distinct distance functions (Section \ref{sec:reg_class}). We then extend this framework by relaxing the deterministic nature of k-means to produce probabilistic regime assignments (Section \ref{sec:prob_dist}). Finally, we show how to derive the regime transition probability matrix within this framework (Section \ref{sec:trans_mat}).

\subsection{Regime Classification}
\label{sec:reg_class}

We use the k-means algorithm with various distance functions to generalize existing strategies used by financial analysts to categorize previous economic periods.
We start by selecting a large set of $m$ macroeconomic variables that represent the market's broader economic state for each month. For each month $t$ being categorized, we have a vector $\boldsymbol{x}_t \in \mathbb{R}^{m}$ that captures the market state at time $t$.

\subsubsection{Separating Outliers using $\ell_2$ Clustering}

The first partition we make of our dataset is to label and separate outlier months. The economic intuition behind such a choice of first step is to identify months with extraordinary market conditions that might not represent the market dynamics in the rest of the dataset. We use the $\ell_2$ distance between every pair of months to perform k-means clustering on our dataset with $k = 2$. Because $\ell_2$ distance is \textit{sensitive to outliers}, this process effectively prunes deviant months and leaves behind the most typical months.

After labeling each month, we are left with two clusters: $A$ and $B$. 
Without loss of generality, we assume $A$ and $B$ are such that $|A| \leq |B|$. In this case, $B$ can be considered the cluster of \textit{typical months} and $A$ the cluster of \textit{outliers}. We begin our regime classification by labeling the months in $A$ as $\textsc{Regime} \; 0$, and turn our attention to the months in $B$ for the remainder of the regime classification algorithm.

\subsubsection{Splitting Typical Samples using Cosine Clustering}

After removing outliers, the months in set $B$ can be considered periods representing ``business as usual''. The $\ell_2$ clustering used to define set $B$ ensures that the state vectors of months within $B$ have magnitudes of similar sizes. To further parse these periods requires switching to a distance metric that is \textit{magnitude agnostic}. Cosine similarity uses the cosine of the angle between two vectors as a measure of distance. It ignores the magnitude of those vectors in its computation, rendering it an apt choice. Therefore, we compute the cosine similarity for each pair of months in $B$, then use a k-means clustering algorithm with $k = r$ to arrive at the final regime classifications for every month in our dataset. We use the simple k-means ``elbow'' heuristic to determine the optimal value of $r$ during cosine clustering.

After both rounds of clustering, we have classified each month into one of the $r + 1$ regimes. $\textsc{Regime} \; 0$ represents outlier months with abnormal macroeconomic conditions, and the remaining regimes in $\{\textsc{Regime} \; 1, ...,  \textsc{Regime} \; r\}$ represent distinct types of regular market behavior.

Algorithm \ref{alg:overview} provides an overview of the regime classification process. This algorithm returns probability distributions over regimes instead of discrete regimes using a mechanism described in Section \ref{sec:prob_dist}. These regime distributions can then be used to improve forecasting models via probabilistic conditioning.

\begin{algorithm}
\caption{An overview of the regime classification process}
\label{alg:overview}
\begin{algorithmic}
\Require $X = \mathbin\Vert_{t=1}^{T} x_t$ ($\mathbin\Vert$ denotes the concatenation operation) is the dataset consisting of the concatenation of all monthly macroeconomic state vectors $x_t$ for $t \in [1, T]$. \\
\State $\{A, B\} \gets \textsc{KMeans}_{[\textsc{$\ell_2$}]}(X, 2)$
\If{$|A| \leq |B|$}
    \State $R_0 \gets A$
    \State $r \gets \textsc{KMeansElbowValue}_{[\textsc{Cosine}]}(B)$
    \State $\{R_1, ..., R_r\} \gets \textsc{KMeans}_{[\textsc{Cosine}]}(B, r)$
\Else
    \State $R_0 \gets B$
    \State $r \gets \textsc{KMeansElbowValue}_{[\textsc{Cosine}]}(A)$
    \State $\{R_1, ..., R_r\} \gets \textsc{KMeans}_{[\textsc{Cosine}]}(A, r)$
\EndIf \\
\Return $\{R_0, ..., R_r\}$
\end{algorithmic}
\end{algorithm}

\subsection{Computing Regime Probability Distributions}
\label{sec:prob_dist}

While the process described so far is intuitive and can produce sufficient regime classifications for every month, to make this tool useful in practice, it should be able to output \textit{probability distributions} for each regime instead of just discrete classifications. This computation is not trivial in our case because of two main factors. To start, the k-means algorithm does not have a direct stochastic algorithm implementation in a cluster label sense. Secondly, two different versions of the k-means algorithm are used consecutively on various subsets of $X$, so any probabilistic outputs must be a combination of the outputs of both instances of the k-means algorithm.

\subsubsection{Probabilistic Distributions from the K-Means Algorithm}

We address the first issue by taking advantage of the fact that, for any given sample to be classified, the k-means algorithm uses the distance between that sample and the cluster centroids of each cluster to classify it. The distances can, therefore, be interpreted as measures of how ``close'' that sample is to being classified as each respective cluster. For a given vector $x$ whose distance from centroid $i$ is $d_i$, we define the probability of that vector being in cluster $C_i$ as

\begin{equation}
    P(C_i) = \frac{1 - \frac{d_i}{\sum_j(d_j)}}{\sum_m \left( 1 - \frac{d_m}{\sum_j(d_j)} \right)}
\end{equation}

We emphasize that this procedure is essentially the same as fuzzy c-mean, although implemented in a slightly different fashion. We use this method to compute distributions for both 
$\textsc{KMeans}_{[\textsc{$\ell_2$}]}$ and $\textsc{KMeans}_{[\textsc{Cosine}]}$.

\subsubsection{Combining Probabilities from $\textsc{KMeans}_{[\textsc{$\ell_2$}]}$ with those from $\textsc{KMeans}_{[\textsc{Cosine}]}$}

Combining the distributions from $\textsc{KMeans}_{[\textsc{$\ell_2$}]}$ and $\textsc{KMeans}_{[\textsc{Cosine}]}$ to get a full distribution over all regimes is not as simple as multiplying the distribution from $\textsc{KMeans}_{[\textsc{Cosine}]}$ by $1 - P(\textsc{Regime} \; 0)$ because doing this can diffuse $1 - P(\textsc{Regime} \; 0)$ too thinly, causing the resulting distribution to be inconsistent with the output of Algorithm \ref{alg:overview}. Instead, we compute the probability distribution of $\{\textsc{Regime} \; 1, ...,  \textsc{Regime} \; r\}$ using the previously described method and then separately calculate a new \textit{scaled value} $P_{R0}$ from $P(\textsc{Regime} \; 0)$.
To this end, we first define
\begin{equation}
    P_{max} = \textsc{max}(\{P(\textsc{Regime} \; 1), ..., P(\textsc{Regime} \; r)\}). 
\end{equation}
We then choose the value for $P_{R0}$ by recognizing that the probability distribution for $\textsc{KMeans}_{[\textsc{$\ell_2$}]}$ can be thought of as a measure of how close $P_{R0}$ should be to $P_{max}$. Before normalizing the distribution to sum to 1, the following must therefore hold
\begin{equation}
    P_{R0} = \begin{cases}
0 & P(\textsc{Regime} \; 0) = 0.0\\
P_{max} & P(\textsc{Regime} \; 0) = 0.5\\
\infty & P(\textsc{Regime} \; 0) = 1.0
\end{cases}
\end{equation}

One such continuous function that would satisfy these conditions cleanly is
\begin{equation}
    P_{R0} = - P_{max} log_{2}(1 - P(\textsc{Regime} \; 0)).
\label{eq:r0}
\end{equation}

We use Equation \eqref{eq:r0} to set $P_{R0}$, then renormalize the distribution to obtain the final regime probability distribution.

\subsection{Computing the Regime Transition Probability Matrix}
\label{sec:trans_mat}

After classifying each month into a regime, computing the transition probability matrix between regimes sheds light on how market dynamics change monthly and describes the relations between regimes.

Each matrix entry can be computed by counting the number of times a given regime switches to every other regime, and dividing that by how many times that regime appears in the dataset. To compute an element $e_{ij}$ of the regime transition probability matrix, we write
\begin{equation}
    e_{ij} = \frac{\textsc{TransitionCount}(\textsc{Regime} \; i, \textsc{Regime} \; j)}{|\textsc{Regime} \; i|}.
\label{eq:transition}
\end{equation}

Figure \ref{fig:transition} depicts an example of such a transition probability matrix.

\section{Regime Analysis}
\label{sec:regime_anal}

Next, we present details of the macroeconomic data we are using (Sections \ref{sec:macro_data}-\ref{sec:data_prep}), compare methods for computing the financial regime associated to each month within the dataset (Section \ref{sec:clus_method_compare}), and present an analysis of those regimes (Section \ref{sec:reg_prop}) along with what they reveal about corresponding market behavior (Sections \ref{sec:market_behave}-\ref{sec:character_reg}).

\subsection{Macroeconomic Data}
\label{sec:macro_data}

In this work, we use 127 variables from the FRED-MD dataset to represent the macroeconomic state of the market. The FRED-MD dataset is a public monthly database of macroeconomic data collected by the Federal Reserve Bank of St. Louis created by \cite{mccracken2016fred} to make data-oriented economic research more accessible. FRED-MD is composed of a variety of groups of variables, namely

\begin{enumerate}
    \item Output and Income
    \item Consumption, Orders, and Inventories
    \item Labor Market
    \item Housing
    \item Money and Credit
    \item Interest and Exchange Rates\label{group}
    \item Prices
\end{enumerate}
We excluded group \ref{group} to focus only on U.S. macroeconomic data.

The variables are sampled monthly and organized into a data frame indexed by timestamps representing the first day of each month. Each row of this data frame then represents the macroeconomic state of the market for that month. This work uses macroeconomic data from December 1959 to January 2023.

\subsection{Data Preparation}
\label{sec:data_prep}

The FRED-MD variables must be standardized before the dataset can be analyzed. To properly perform this standardization, most of the variables in the FRED-MD dataset must be transformed before they can be demeaned and standardized to account for their different arrangements (indices, prices, rates, etc.). The appropriate transformations to apply before standardization are pre-defined by the Federal Bank of St. Louis (each variable has a \textit{t-code} assigned to it that corresponds to one of seven possible transformations).

\begin{figure}[ht]
    \centering
    \includegraphics[width=0.45\textwidth]{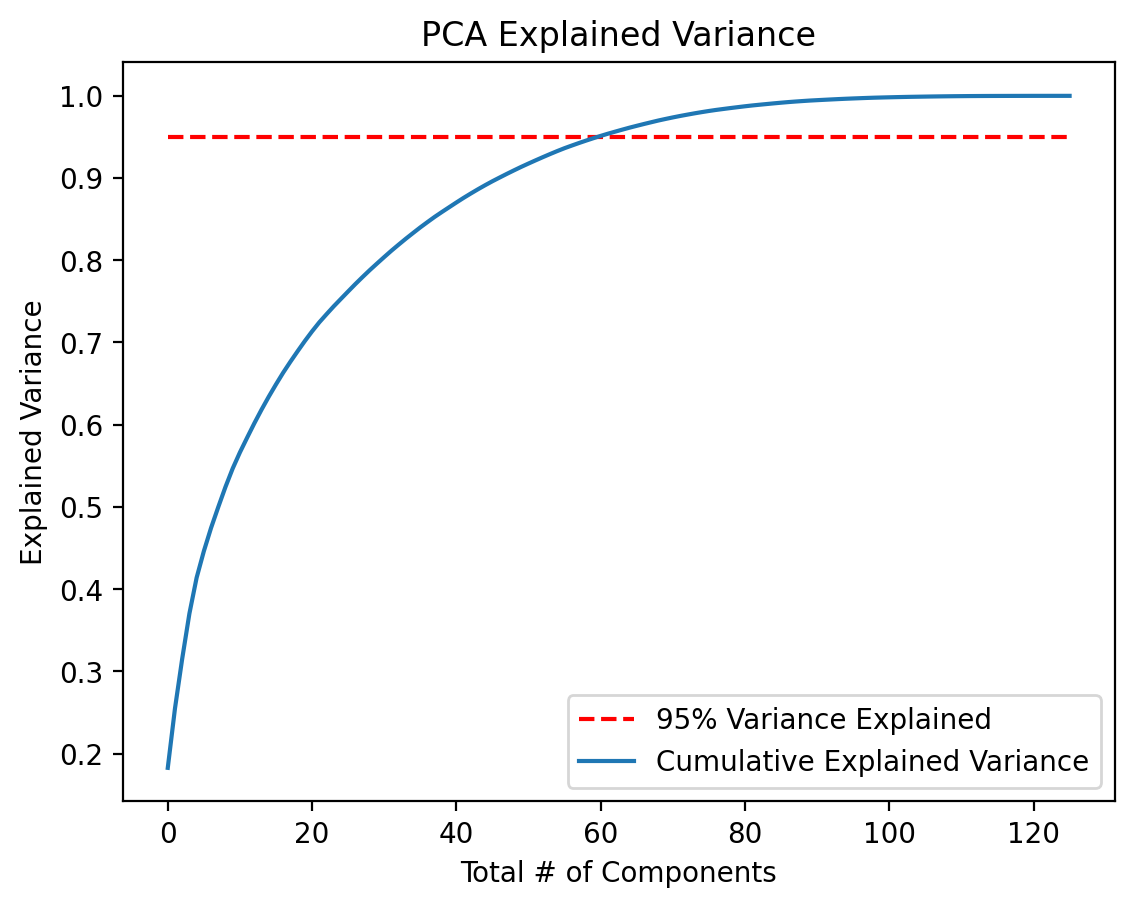}
    \caption{The cumulative variance explained from using different amounts of eigenvector components in the PCA dimensionality reduction process. The curve reaches 95\% at around the 61-component mark.}
    \label{fig:pca_expl}
\end{figure}

We apply these transformations, standardize the variables, and then use principle component analysis (PCA) to find the eigenvectors of the data with the largest corresponding eigenvalues. These are used to reduce the number of dimensions of the dataset and extract the most critical factors. The amount of variance explained by each component can be cumulatively summed in descending order of their eigenvalues until a preset variance explanation threshold is met. In this work, we use a threshold of 95\%, corresponding to 61 components. A visualization of how much variance is explained by varying the number of components is shown in Figure \ref{fig:pca_expl}.

We also set the $k$ value of our algorithm's second k-means clustering layer using the previously mentioned k-means ``elbow'' heuristic. The optimal value of $k$ to use in the $\textsc{KMeans}_{[\textsc{Cosine}]}$ part of our methodology is $k = 5$. We, therefore, set $k$ to this value in this work.

\subsection{Clustering Method Comparison}
\label{sec:clus_method_compare}

This section compares the modified k-means method with the Gaussian Mixture Model (GMM) to illustrate the differences between an algorithmic clustering approach and a model-based one. While the GMM is inherently probabilistic and efficiently extracts probabilities for each regime assignment, the modified k-means relies on a heuristic to estimate probabilities, offering a different perspective on cluster confidence and transitions.

\begin{figure}[ht] 
\centering \includegraphics[width=0.99\textwidth]{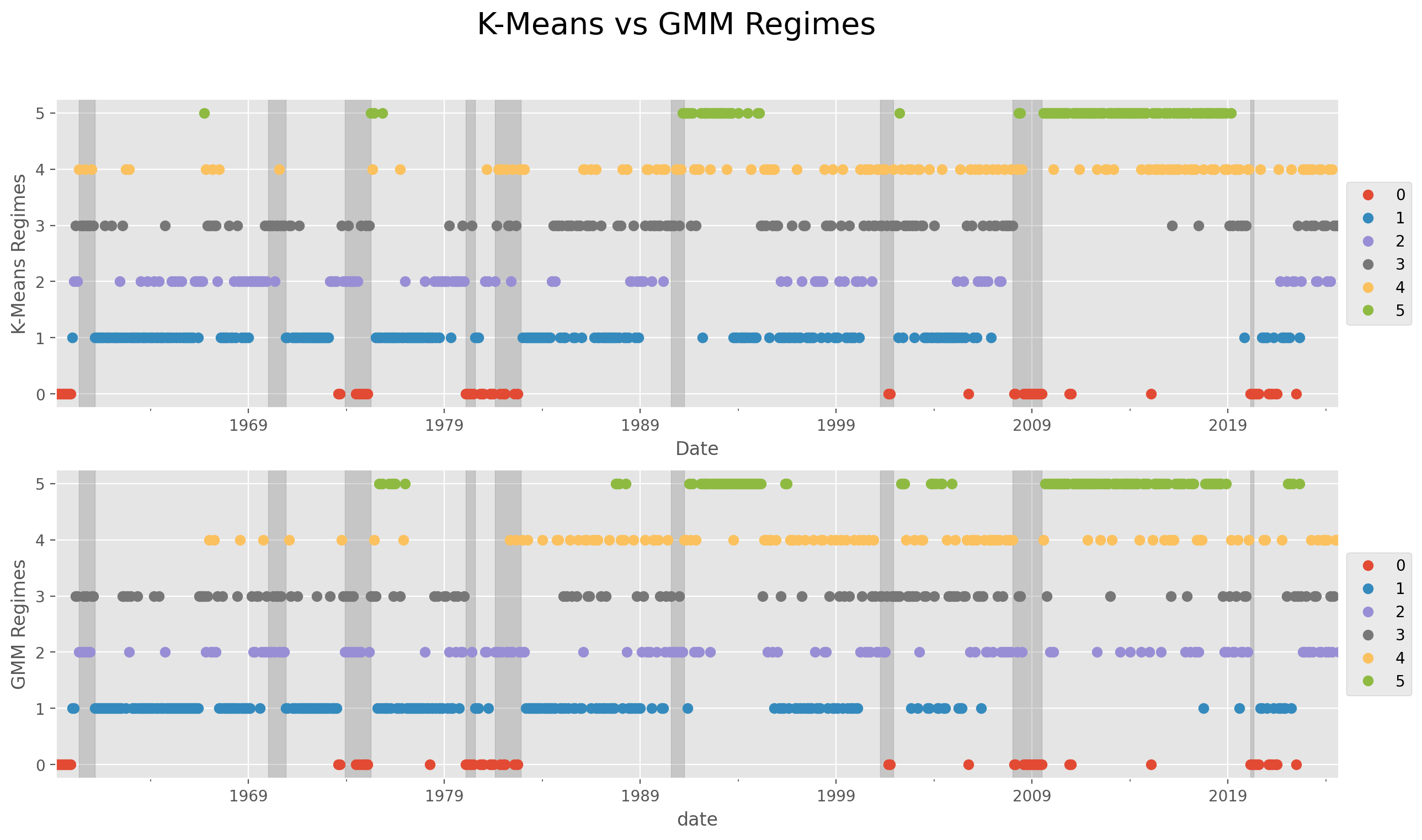} 
\caption{A comparison of the regimes detected using the layered k-means method and the analogous GMM method. NBER recession periods are highlighted, alongside regime labels for both methods. Most classifications align, with notable differences between the two methods observed in $\textsc{Regime} \; 2$ and $\textsc{Regime} \; 3$.}
\label{fig:gmm_compare} 
\end{figure}

Figure \ref{fig:gmm_compare} compares the regime labels assigned by the modified k-means and GMM methods. The gray-shaded areas denote NBER-defined recession periods. Both methods capture similar regimes, aligning on most classifications, with only minor differences observed in transitions, particularly for regimes 2 and 3. This suggests that while both methods are robust in identifying consistent patterns, their unique sensitivities to data structure lead to slight variations. Both methods also strongly align with the NBER-defined recession periods, confirming their ability to detect macroeconomic shifts effectively.

\begin{figure}[ht] 
\centering 
\includegraphics[width=0.99\textwidth]{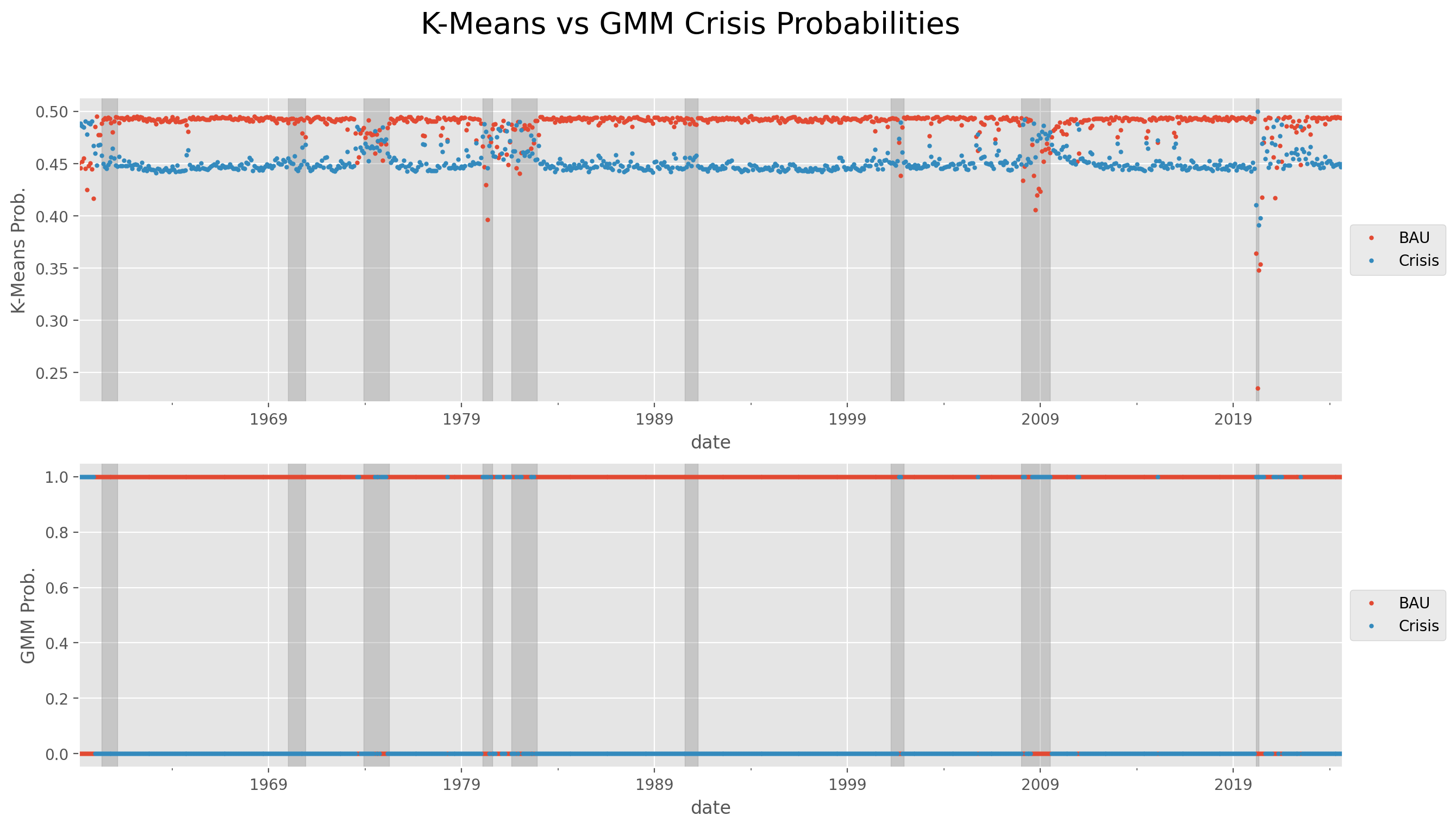}
\caption{A comparison of the probabilities of being in a crisis period versus a normal period, as computed using the k-means distribution heuristic and the GMM. NBER recession periods are highlighted. The k-means probabilities show gradual transitions, while GMM distributions appear deterministic.}
\label{fig:gmm_crisis}
\end{figure}

Figure \ref{fig:gmm_crisis} compares crisis probabilities derived from the modified k-means heuristic and the GMM. While the k-means probabilities exhibit smoother transitions over time, suggesting a more nuanced approach to modeling uncertainty, the GMM assigns almost binary probabilities that reflect strong confidence in its cluster assignments. This probabilistic framework of the GMM ensures clarity in classification but may overlook subtle transitions. On the other hand, the gradual shifts in the k-means probabilities highlight its flexibility in capturing heterogeneity within clusters, where it adjusts probabilities to reflect varying levels of certainty as new data becomes available.

Interestingly, both methods align on specific probability spikes, indicating agreement on key transition points, particularly during significant economic turning points. This alignment underscores the reliability of these clustering approaches in identifying critical periods of economic shifts. However, the modified k-means method offers a more interpretable view of uncertainty, rendering it particularly valuable for applications such as portfolio allocation, where understanding gradual transitions rather than abrupt changes can guide better risk management and decision-making strategies.

Overall, this comparison highlights the trade-offs between the two approaches. The GMM provides high-confidence regime classifications that are nearly deterministic, making it suitable for applications requiring clear, binary outcomes. In contrast, the k-means approach balances flexibility and interpretability, capturing the gradual dynamics of regime transitions. Both methods show strong consistency with NBER-defined recession periods, demonstrating their ability to detect macroeconomic shifts effectively. However, the probabilistic nuance of the k-means method provides an additional layer of insight into regime dynamics, which can be critical for understanding the complexities of financial markets.

\subsection{Regime Properties}
\label{sec:reg_prop}

\begin{figure}
    \centering
    \includegraphics[width=0.50\textwidth]{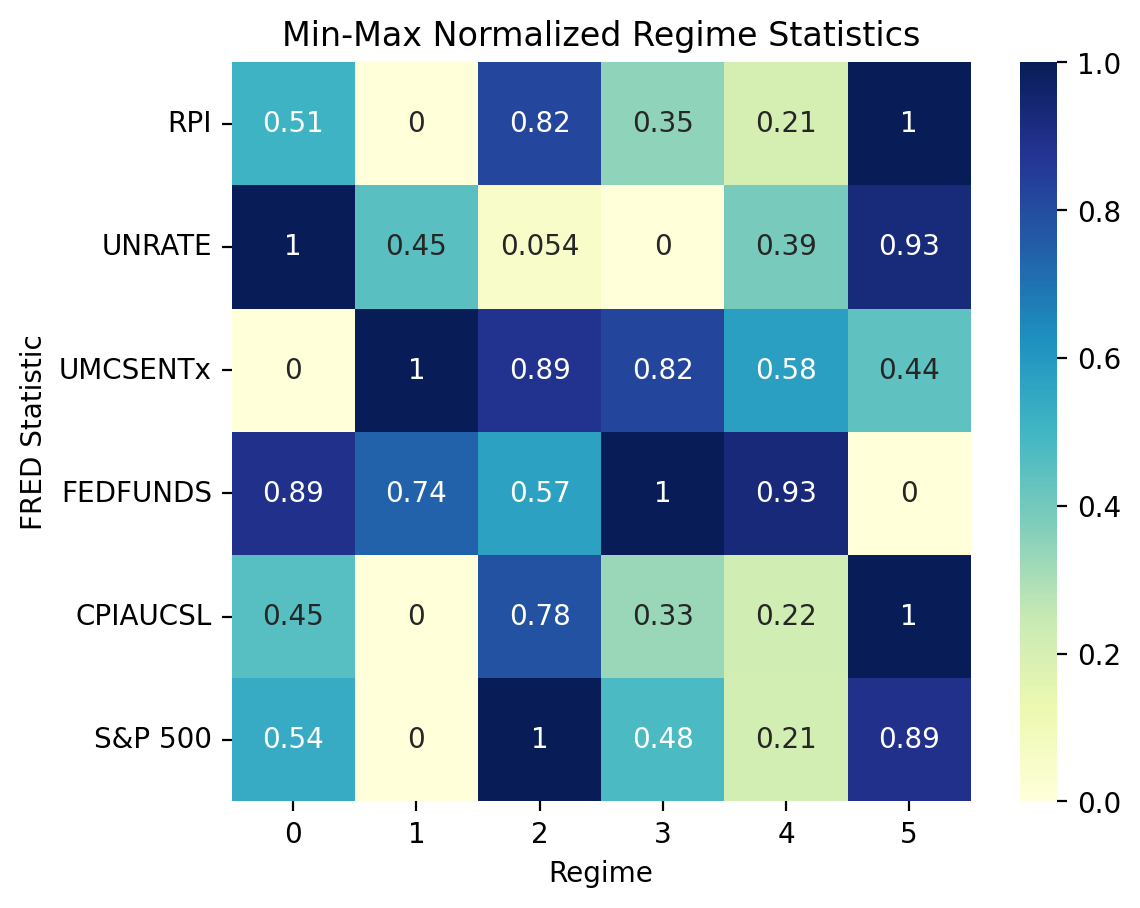}
    \caption{Macroeconomic differences across regimes as measured via the average value of crucial FRED-MD statistics. The value of the statistics are min-max normalized along each row to make them easier to compare with one another.}
    \label{fig:reg_stats}
\end{figure}

One can understand the economic differences between regimes by measuring the average value of certain crucial FRED-MD statistics within each regime. Figure \ref{fig:reg_stats} provides this type of analysis.
Here, we observe that $\textsc{Regime} \; 0$, the regime containing all of the ``outliers'' months filtered out using the $\textsc{KMeans}_{[\textsc{$\ell_2$}]}$ algorithm, best corresponds to periods of economic difficulty. To support this intuition, $\textsc{Regime} \; 0$ has the lowest real personal income, consumer sentiment, federal funds rate, and the highest unemployment rate. This implies that the effect of stacking $\textsc{KMeans}_{[\textsc{$\ell_2$}]}$ and $\textsc{KMeans}_{[\textsc{Cosine}]}$ in the regime labeling process is first to separate the atypical periods of economic panic, then cluster the remaining months of normal economic activity based on the direction of their macroeconomic state vectors.

\subsection{Market Behavioral Analysis Using Regimes}
\label{sec:market_behave}

\begin{figure}
    \centering
    \includegraphics[width=0.45\textwidth]{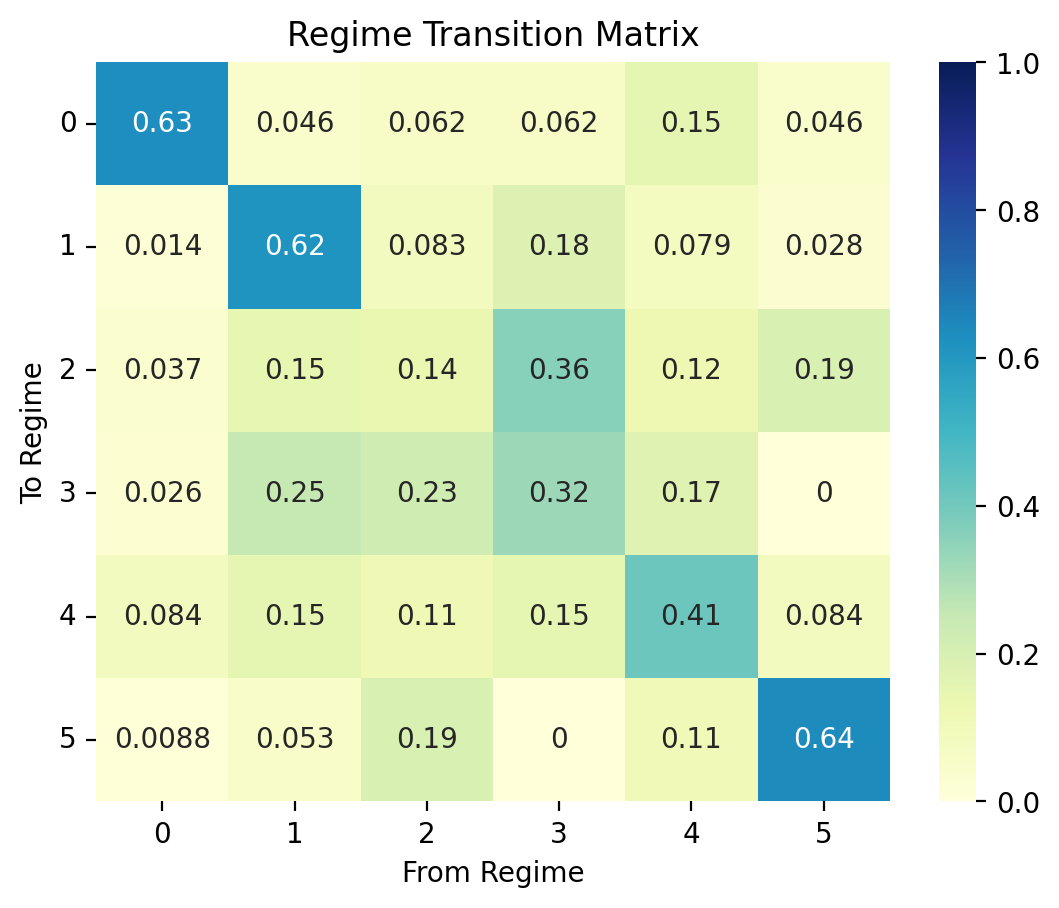}
    \includegraphics[width=0.45\textwidth]{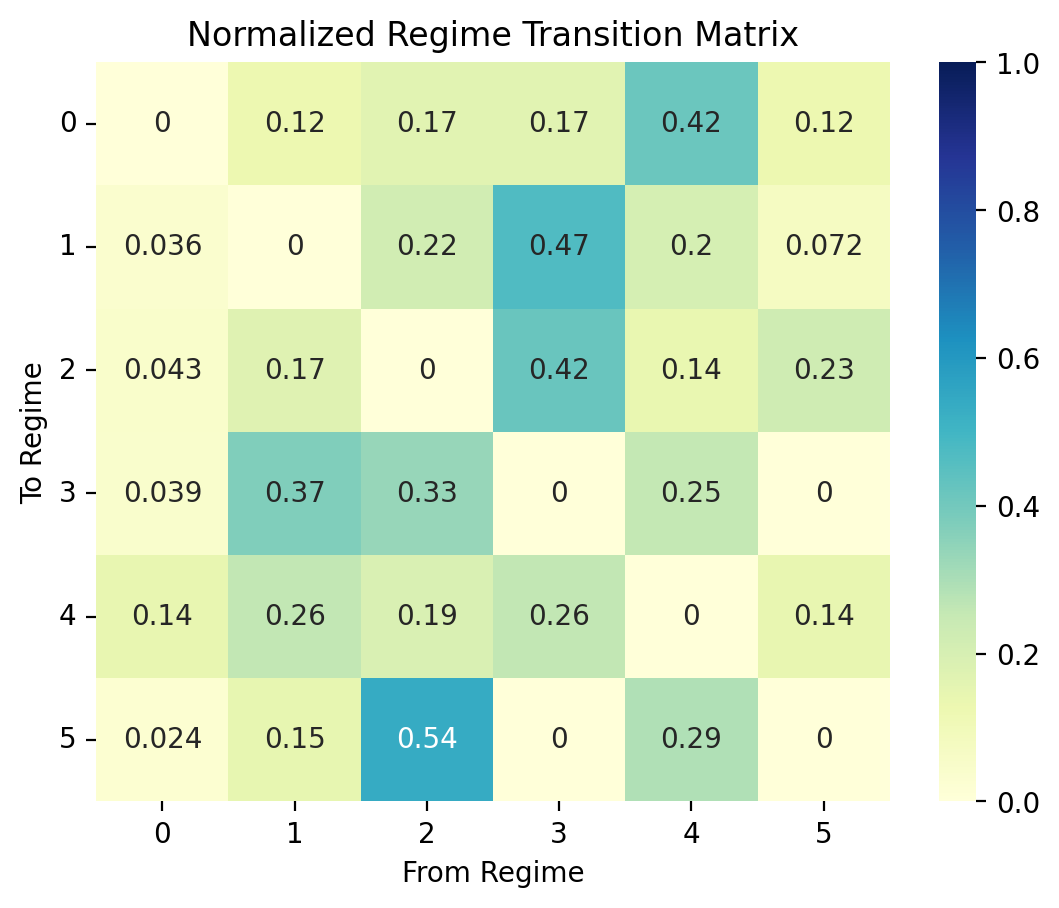}
    \caption{\textit{Left:} Regime transition probability matrix computed using all available data. \textit{Right:} Regime transition probability matrix given that a regime transition occurs.}
    \label{fig:transition}
\end{figure}

Using the regime transition probability matrix computed from the monthly regime timeline, we are now able to learn how regimes are related and how a market will likely change over time. We present such a matrix in Figure \ref{fig:transition}, and draw the following conclusions. 

We observe that most of the probability density lies on the diagonal. This shows us that the most likely regime of the next month is always the regime of the current month, i.e., regime shifts do not occur with high frequency, which is what we expect to observe given the existing literature on macroeconomic regimes such as \cite{baele2015macroeconomic}.

We also note that some matrix elements have shallow values, with some being precisely 0. This implies that specific regime transitions are impossible or at least extremely unlikely, which is also what we would expect to see in a well-functioning macroeconomic regime detection model.

\subsection{Characterizing Regimes}
\label{sec:character_reg}

We characterize and label each regime descriptively using the information in Figures \ref{fig:reg_stats} and \ref{fig:transition}. While these characterizations cannot be absolute because of the approximate nature of the k-means algorithm, we present our regime descriptions in Table \ref{tab:reg_desc}.

\begin{table*}[ht]
    \centering
    \caption{Economic characterizations of each of the regimes in this work.}
    \begin{tabular}{llp{9.5cm}}\toprule
    Regime Index & Regime Label & Characteristic Description \\\midrule
    0 & Economic Difficulty & This regime has the highest unemployment rate and lowest consumer sentiment, federal funds rate, and real personal income. This regime is also the most self-stable and likely to transition to stagflation.\\
    1 & Economic Recovery & This regime represents a recovery phase where inflation is controlled, but equity markets lag due to uncertainty. High consumer sentiment here suggests there is public confidence in future growth.\\
    2 & Expansionary Growth & This regime has strong economic prosperity. Good equity market performance and high consumer sentiment reflect robust economic confidence. Inflation is moderate, and unemployment is balanced, creating favorable conditions for growth. Central banks typically maintain a neutral monetary policy.\\
    3 & Stagflationary Pressure & This regime is marked by high inflation and rising interest rates. While unemployment may not be extreme, the restrictive monetary policy dampens equity markets. This regime aligns with stagflationary environments where growth stagnates alongside rising prices.\\
    4 & Pre-Recession Transition & This regime could be a warning sign before a recession. Inflation begins to cool, but economic activity slows, leading to subdued equity markets. Unemployment rises slightly, and monetary policy starts tightening to combat earlier inflation, which can trigger economic downturns.\\
    5 & Reflationary Boom & This regime reflects periods of economic resurgence fueled by loose monetary policy and strong equity market performance. Inflation is high, but central banks prioritize growth and employment over price stability. These periods often coincide with quantitative easing or other accommodative measures.\\\bottomrule
    \end{tabular}
    \label{tab:reg_desc}
\end{table*}

To help us better understand these regimes, we further modify the transition probability matrix in Figure \ref{fig:transition} to represent transition probabilities \textit{given that a transition takes place}. This can be achieved by dividing all the off-diagonal elements of row $i$ by $1 - \textsc{diag}(i)$, where $\textsc{diag}(i)$ is the $i$-th diagonal element of the transition probability matrix, then setting all the diagonal elements to 0. After this step, we arrive at the normalized matrix in Figure \ref{fig:transition}.

Using this new matrix, we now define a graph structure over the regimes using the matrix from Figure \ref{fig:transition} as the graph adjacency matrix. Here, the edge weights are characterized by the probability that one regime will transition to another, given that a transition occurs.

\begin{figure}
    \centering
    \includegraphics[width=0.65\textwidth]{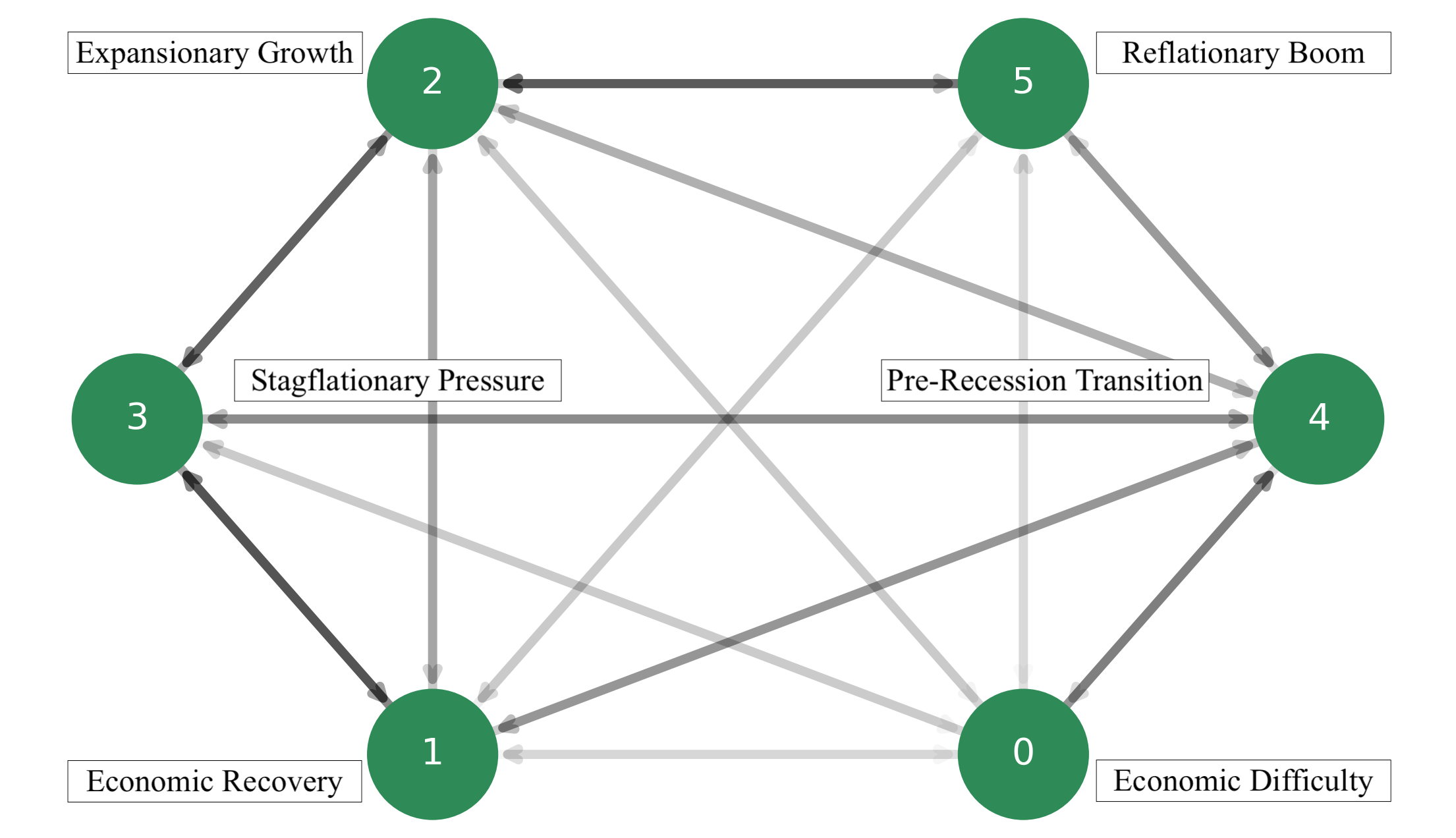}
    \caption{Regime transitions visualized in network form. Darker edges signify larger transition probabilities, and nodes have been positioned to emphasize the emergent path graph between the regimes. Labels are taken from Table \ref{tab:reg_desc}.}
    \label{fig:reg_graph}
\end{figure}

The graph from this procedure, shown in Figure \ref{fig:reg_graph}, helps better understand how the market changes over time. The edges are colored according to their edge weight, and the positions of the nodes have been rearranged to highlight the strong path graph that emerges from the regime data.

\section{Forecasting Experiments}

We now demonstrate how these regimes can be leveraged to augment forecasting models to improve their forecasting accuracy.
For this work, we chose to focus on a tactical asset allocation task for a portfolio made up of only ETFs and index funds, because we anticipate these are the types of assets that will be affected the most  by the market's macroeconomic conditions.

\subsection{Regimes as Probabilistic Conditions}

Regimes can enhance forecasting models by serving as a probabilistic condition on the forecast of the underlying asset. Ideally, we should consider the current regime estimate and the regime transition probability matrix representing the most probable following regimes. It is essential to acknowledge the uncertainty surrounding these estimates. When dealing with the uncertainty around current regimes, instead of treating the output label for the current month as a deterministic fact, we should consider the distribution of possible current regimes.

In practice, this means that we should combine the distribution over the current regimes and the regime transition probability matrix before applying it to any forecasting model of choice. To describe this procedure more precisely, recall that $P(\textsc{Regime} \; i)$ is the probability measure associated with the regimes $i$, for $i = 0, ..., r$, induced by the procedure described in Section \ref{sec:prob_dist}. Let $\boldsymbol{p}_t \in \mathbb{R}^{r + 1}_{+}$ be the discrete probability values associated with each regime for the $t$-th time step, and let $E_t \in \mathbb{R}^{(r + 1) \times (r + 1)}_{+}$ denote the regime transition probability matrix defined in Section \ref{sec:trans_mat}.

As mentioned at the end of Section \ref{sec:prob_dist}, we first need to ensure that $\boldsymbol{p}_t$ is renormalized to sum to 1. We achieve  this by dividing it by its magnitude as follows
\begin{equation}
    \Tilde{\boldsymbol{p}}_t = \frac{\boldsymbol{p}_t}{|\boldsymbol{p}_t|}. 
\end{equation}

Second, we update the current state probability by combining both the distribution over current regimes and the transition matrix 
\begin{equation}
    \Tilde{\boldsymbol{p}}_{t+1} = \Tilde{\boldsymbol{p}}^{\top}_t  E_t.
    \label{eq:prob_update}
\end{equation}

Indeed, this process is the same as when iterating a Markov chain process \citep{ross-2019}. It is important to note that, as a less effective alternative, we could fix the current state to be the one with the highest probability. However, this process would completely ignore the uncertainty around the current state estimates and would result in a model whose probabilistic conditioning would be identical for months in the same regime despite their macroeconomic state vectors being different.

\subsection{Forecasting Models}
\label{sec:models}

We employ three simple forecasting strategies to demonstrate how knowledge of the current market conditions and transition probabilities can improve model accuracy. The first strategy creates predictions using past Sharpe ratios conditioned on the most probable regime. The second strategy uses the framework proposed by \cite{black-litterman-1991}, where sample mean and covariance returns are combined with priors based on our proposed regime model. Finally, the last strategy builds specialized models by estimating parameters for each market regime and employs these to make predictions. The final predictions are then adjusted based on the updated estimates of the current market condition probabilities, as defined in Equation \eqref{eq:prob_update}.

Let $\hat{y}^{j}_{t+1}$ represent the prediction for time step $t+1$ made by any of the considered models for ETF $j$. Let $s^{j}_{1:t}$ denote the Sharpe ratio computed using returns from $1$ to $t$. Lastly, let $\boldsymbol{X}^{i}_{1:t}$ be a matrix of macroeconomic features observed on $\textsc{Regime} \; i$ from $1$ to $t$. Next, we will define each of the forecast models using this notation.

\subsubsection{Naive Forecasting Model}
The naive forecasting model is a simple forecasting scheme designed to evaluate the effectiveness of the regime module. It is parameter-free and relies solely on historical data conditional on regime classifications.

For each time step $t+1$, we first identify the most likely regime $i^{*}_{t+1}$, defined as
\begin{equation}
   i^{*}_{t+1} = \underset{i \in \{ 1,...,r+1 \} }{\operatorname{argmax}} \; \Tilde{p}_{i, t+1}.
\end{equation}

Given this predicted regime, the forecast for ETF $j$ at time step $t+1$ is the expected Sharpe ratio conditional on the regime
\begin{equation}
    \hat{y}^{j, \text{naive}}_{t+1} = \mathbb{E}\left[s_{j, t+1}|\textsc{Regime} = i^{*}_{t+1}\right] = \hat{s}^{*}_{j,1:t},
    \label{eq:eq9}
\end{equation}
\noindent where $\hat{s}^{*}_{j,1:t}$ is the sample estimate of the conditional Sharpe ratio
\begin{equation}
     \hat{s}^{*}_{j,1:t} = \frac{\hat{\mu}^{*}_{j,1:t}}{\hat{\sigma}^{*}_{j,1:t}}
\end{equation}

Here, $\hat{\mu}^{*}_{j,1:t}$ and $\hat{\sigma}^{*}_{j,1:t}$ are the sample mean and standard deviation, respectively, calculated using data from time steps 1 to $t$ where the regime matches $i^{*}_{t+1}$. Thus, the estimation in Equation \eqref{eq:eq9} uses only the subset of historical data corresponding to the predicted regime.



\subsubsection{Black-Litterman Model}
The model proposed by \cite{black-litterman-1991}  combines market equilibrium returns with investors' own views to generate adjusted expected returns. In our implementation, instead of using market equilibrium returns derived from betas, we use sample means and covariances as prior beliefs. These are then updated with investor views which in our case are constructed from regime-conditional expected returns.

For each ETF $j$, the sample mean $\mu_j$ gives the prior expected return, with uncertainty captured by the sample covariance matrix $\Sigma$. Using our regime model, the views are derived from the expected returns conditional on the most likely next regime $i^*_{t+1}$.

For each ETF $j$, the view on its expected return is
\begin{equation}
   q^{*}_{j, t+1} = \hat{y}^{j, \text{bl}}_{t+1} = \mathbb{E}\left[r^{j}_{t+1}|\textsc{Regime} = i^*_{t+1}\right] = \hat{\mu}^{*}_{j,1:t}
\end{equation}
\noindent 
where $\hat{\mu}^{*}_{j,1:t}$ is the sample mean return calculated using data from time steps 1 to $t$ where the regime matches $i^*_{t+1}$. These $q_{j, t+1}$ will later be used for building allocations in the Black-Litterman framework. In particular, these values are used as the ``views'' in the Black-Litterman model.



\subsubsection{Linear Ridge Regression Model}
The linear ridge model proposed by \cite{hoerl-1968, hoerl-1970} is a step further using simple learning models to predict the next time step returns. There are two main differences between this model and the naive benchmark: (1) it predicts returns instead of Sharpe ratios, and (2) it has learnable parameters that get updated with each new time step.

Following up the intuition built above, the forecast of the ridge model for each $\textsc{Regime} \; i$ and ETF $j$ is given by
\begin{equation}
    \hat{y}^{i, j, \text{ridge}}_{t+1} = \mathbb{E}\left[r^{j}_{t+1}|\textsc{Regime} \; i \right] = \hat{\boldsymbol{\beta}}^{i, j}_{\text{ridge}}\boldsymbol{X}^{i}_{1:t}, 
\end{equation}
where
\begin{equation}\hat{\boldsymbol{\beta}}^{i,j}_{\text{ridge}} = \underset{\boldsymbol{\beta}^{i, j}}{\operatorname{argmin}} \left\{ (\boldsymbol{y}^{j}_{1:t} - \boldsymbol{X}^{i}_{1:t}\boldsymbol{\beta}^{i, j})^\top (\boldsymbol{y}^{j}_{1:t} - \boldsymbol{X}^{i}_{1:t}\boldsymbol{\beta}^{i, j}) + \lambda \|\boldsymbol{\beta}^{i, j}\|_2^2 \right\}.
\end{equation}

Once we have available all the forecasts for each ETF and the regime, we again use the values in $\Tilde{\boldsymbol{p}}_{t+1}$ to aggregate the forecasts through the regime dimension such that we end up with a final probability weighted forecasts for each ETF, as follows
\begin{equation}
    \hat{y}^{j, \text{ridge}}_{t+1} = \sum_{i=1}^{r} \Tilde{p}_{i, t+1} \cdot \hat{y}^{i, j, \text{ridge}}_{t+1}.
\end{equation}

\subsection{Position Sizing}
\label{sec:pos_sizing}

\subsubsection{Standard Position Sizing}
Let $\hat{\boldsymbol{y}}^{k}_{t+1} \in \mathbb{R}^{d}$, for $k \in \{\text{naive}, \text{ridge} \}$ be a vector of predictions for all $d$ ETFs. Let also $\mathcal{H}$ and $\mathcal{L}$ be two sets composed of the $l$ highest and $l$ lowest values of $\hat{\boldsymbol{y}}^{k}_{t+1}$ respectively. We assume that $\mathcal{H}$ has only positive values and scale the size of $\mathcal{H}$ down accordingly if this is not the case. We follow similar steps for negative numbers with $\mathcal{L}$. Finally, let $g: \mathbb{R}^{d} \rightarrow [-1, 1]^d$ be a function that maps returns into position sizing allocations.

We begin by defining $S$, the magnitude of our asset choices as
\begin{equation}
    S_l = \sum_{\hat{y}^{j, k}_{t+1} \in \mathcal{H} \cup \mathcal{L}} | \hat{y}^{j, k}_{t+1} |.
\end{equation}
Next, we define a ``long and short'' (lns) position sizing strategy as
\begin{equation}
    \boldsymbol{w}^{\text{lns}}_{t+1} = g_{\text{lns}}(\hat{\boldsymbol{y}}^{k}_{t+1}, l) = 
    \begin{cases} 
        \frac{\hat{y}^{j, k}_{t+1} }{S_l} & \text{for } \hat{y}^{j, k}_{t+1} \in \mathcal{H} \cup \mathcal{L}\\\
        0 & \text{otherwise}.
    \end{cases}
\end{equation}

A long and short strategy has the drawback of requiring us to take both a short and long position regardless of the magnitude of the forecasts. Instead, we can define a ``long or short'' (los) strategy that only takes positions on the $l$ ETFs that have forecasts with the \textit{largest magnitude} as
\begin{equation}
    \boldsymbol{w}^{\text{los}}_{t+1} = g_{\text{los}}(\hat{\boldsymbol{y}}^{k}_{t+1}, l) = 
    \begin{cases} 
        \frac{\hat{y}^{j, k}_{t+1} }{S_l} & \text{for } \hat{y}^{j, k}_{t+1} \in \mathcal{B}\\\
        0 & \text{otherwise}, 
    \end{cases}
\end{equation}
where $\mathcal{B}$ is the set of ETFs whose forecasts have the largest magnitude, and $S_l$ is redefined to use $\mathcal{B}$ instead of $\mathcal{H} \cup \mathcal{L}$ in its summation.

Suppose instead, we seek to eliminate drawdowns from short positions. In that case, we may redefine $g$ appropriately to use a ``long-only instead'' (lo) position sizing strategy by recomputing $S_l$ to use only $\mathcal{H}$ instead of $\mathcal{H} \cup \mathcal{L}$ in its summation. It would have allocations
\begin{equation}
    \boldsymbol{w}^{\text{lo}}_{t+1} = g_{\text{lo}}(\hat{\boldsymbol{y}}^{k}_{t+1}, l) = 
    \begin{cases} 
        \frac{\hat{y}^{j, k}_{t+1}}{S_l} & \text{for } \hat{y}^{j, k}_{t+1} \in \mathcal{H}\\
        0 & \text{otherwise}. 
    \end{cases}
\end{equation}

As we will see in Section \ref{sec:res}, we find that using a long-only strategy is optimal except for periods of economic difficulty when shorts yield substantial returns. To account for this finding, we define one final position sizing strategy that mixes both the long-only strategy and the long or short strategy. It uses a long or short strategy if the following month is forecasted to be in $\textsc{Regime} \; 0$ (the regime associated with economic difficulty), and a long-only strategy for every other regime. This strategy is called the ``mixed'' (mx) strategy.

\subsubsection{Black-Litterman Position Sizing}
Given the regime-conditional views $q_{j,t+1}$ for each ETF $j$, we can construct the portfolio weights using the Black-Litterman framework. The posterior distribution of expected returns leads to the following allocation
\begin{equation}
 \boldsymbol{w}_{t+1} = [(\tau\Sigma_{1:t})^{-1} + P_{t+1}^\top\Omega_{t+1}^{-1}P_{t+1}]^{-1}[(\tau\Sigma_{1:t})^{-1}\hat{\boldsymbol{\mu}}_{1:t} + P_{t+1}^\top\Omega_{t+1}^{-1}\boldsymbol{q}^{*}_{t+1}],
\end{equation}
\noindent  where $\hat{\boldsymbol{\mu}}_{1:t}$ and $\Sigma_{1:t}$ are the sample mean vector and covariance matrix estimated using data from time steps 1 to $t$, $\tau$ is a scaling parameter representing the uncertainty in the prior beliefs, $P_{t+1}$ is the pick matrix expressing which assets have views for time $t+1$, $\boldsymbol{q}^{*}_{t+1}$ is the vector of regime-conditional expected returns derived in the previous section, and $\Omega_{t+1}$ represents the uncertainty matrix for our regime-conditional views.

\subsection{Performance Metrics}

The performance of different portfolio models was evaluated using several metrics: Sharpe ratio, Sortino ratio, average drawdown (AvgDD), maximum drawdown (MaxDD), and the percentage of positive returns (\% Positive Ret.). We present results for four different models: the naive model (Naive), linear ridge model (Ridge), mean-variance optimization model (MVO), and Black-Litterman model (BL). For each model, we analyze four types of portfolios: long-only (lo), long and short (lns), long or short (los), and mixed (mx) portfolios, as described in Section \ref{sec:pos_sizing}.

\section{Experimental Results}
\label{sec:res}

We begin our experimental setup with the FRED-MD dataset, which spans from February 2000 to January 2023, as detailed in Section \ref{sec:regime_anal}. This macroeconomic dataset is merged with beginning-of-the-month returns for ten Exchange-Traded Funds (ETFs), described in Table \ref{tab:etf_desc}, sourced from the Wharton Research Data Services (WRDS). The ETF dataset covers the period from February 2000 to December 2022.

After merging the two datasets, the final dataset comprises 746 monthly observations from 2000 to 2022. For our analysis, we define an estimation window of 48 months (four years) for both the regime detection and prediction models. Using this estimation window, we implement a fixed-window, walk-forward prediction exercise, where predictions are made one month ahead. We use these predictions to build portfolios as defined in Section \ref{sec:pos_sizing}.

{\scriptsize
\begin{table}[h]
    \centering
    \begin{tabular}{@{}lll@{}}
        \toprule
        ETF Ticker & Sector & Description \\
        \midrule
        SPY & S\&P 500 & Tracks the largest 500 U.S. companies. \\
        XLB & Materials & Covers chemicals, mining, and construction materials. \\
        XLE & Energy & Includes oil, gas, and renewable energy companies. \\
        XLF & Financials & Banks, insurers, and asset managers. \\
        XLI & Industrials & Aerospace, defense, and manufacturing. \\
        XLK & Technology & Software, hardware, and IT services. \\
        XLP & Staples & Essential goods like food and household products. \\
        XLU & Utilities & Electricity, water, and gas providers. \\
        XLV & Healthcare & Pharmaceuticals, biotech, and providers. \\
        XLY & Discretionary & Retail, automotive, and entertainment. \\
        \bottomrule
    \end{tabular}
    \caption{Economic characterizations of ETFs used in this work.}
    \label{tab:etf_desc}
\end{table}
}

\begin{figure}[H]
    \centering
    \includegraphics[width=0.8\textwidth]{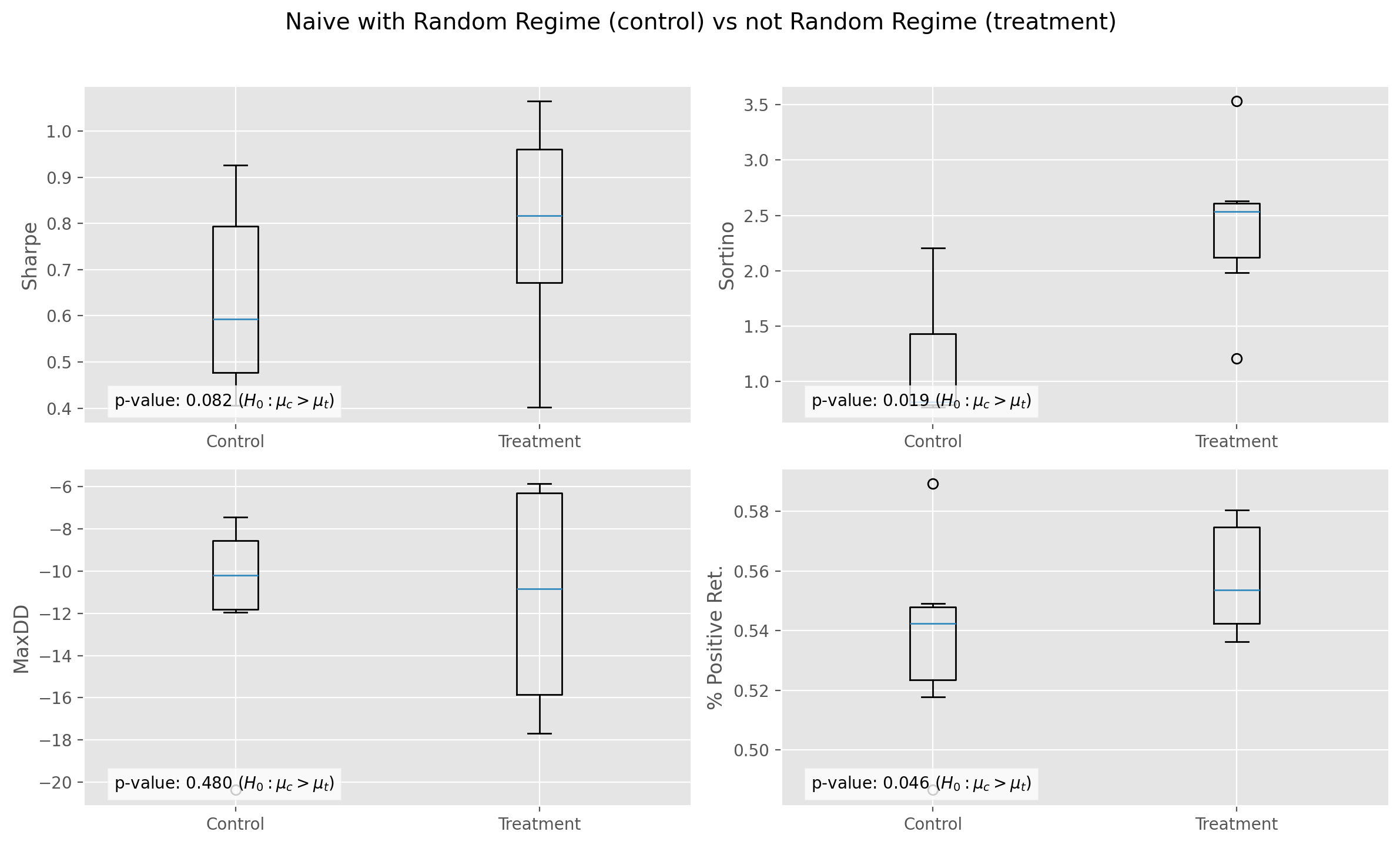}
    \caption{The figure depicts boxplots for controls and treatment portfolio metrics. The control is defined as the the naive method based on random regimes, whereas the control is defined as the naive method based on non-random regimes. The figure also shows p-values for the t-test for two related samples where the null hypothesis is that the control mean is higher than the treatment mean.}
    \label{fig:fig7}
\end{figure}

\begin{figure}[H]
    \centering
    \includegraphics[width=0.8\textwidth]{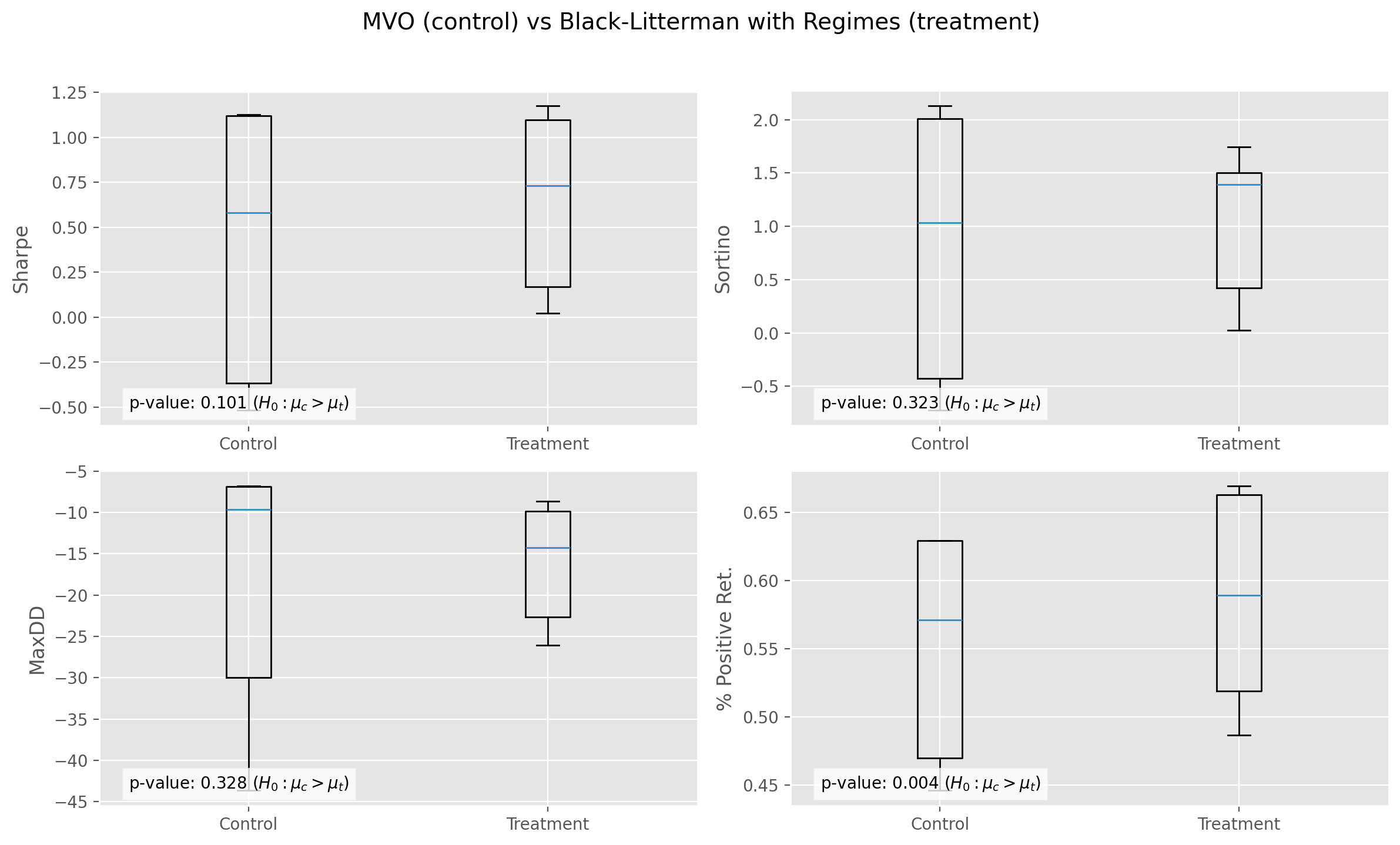}
    \caption{The figure depicts boxplots for controls and treatment portfolio metrics. The control is defined as the the mean-variance optimization method (mvo), whereas the control is defined s the black-litterman (bl) where the views are defined based on the regimes. The figure also shows p-values for the t-test for two related samples where the null hypothesis is that the control mean is higher than the treatment mean.}
    \label{fig:fig8}
\end{figure}

\begin{figure}[H]
    \centering
    \includegraphics[width=0.8\textwidth]{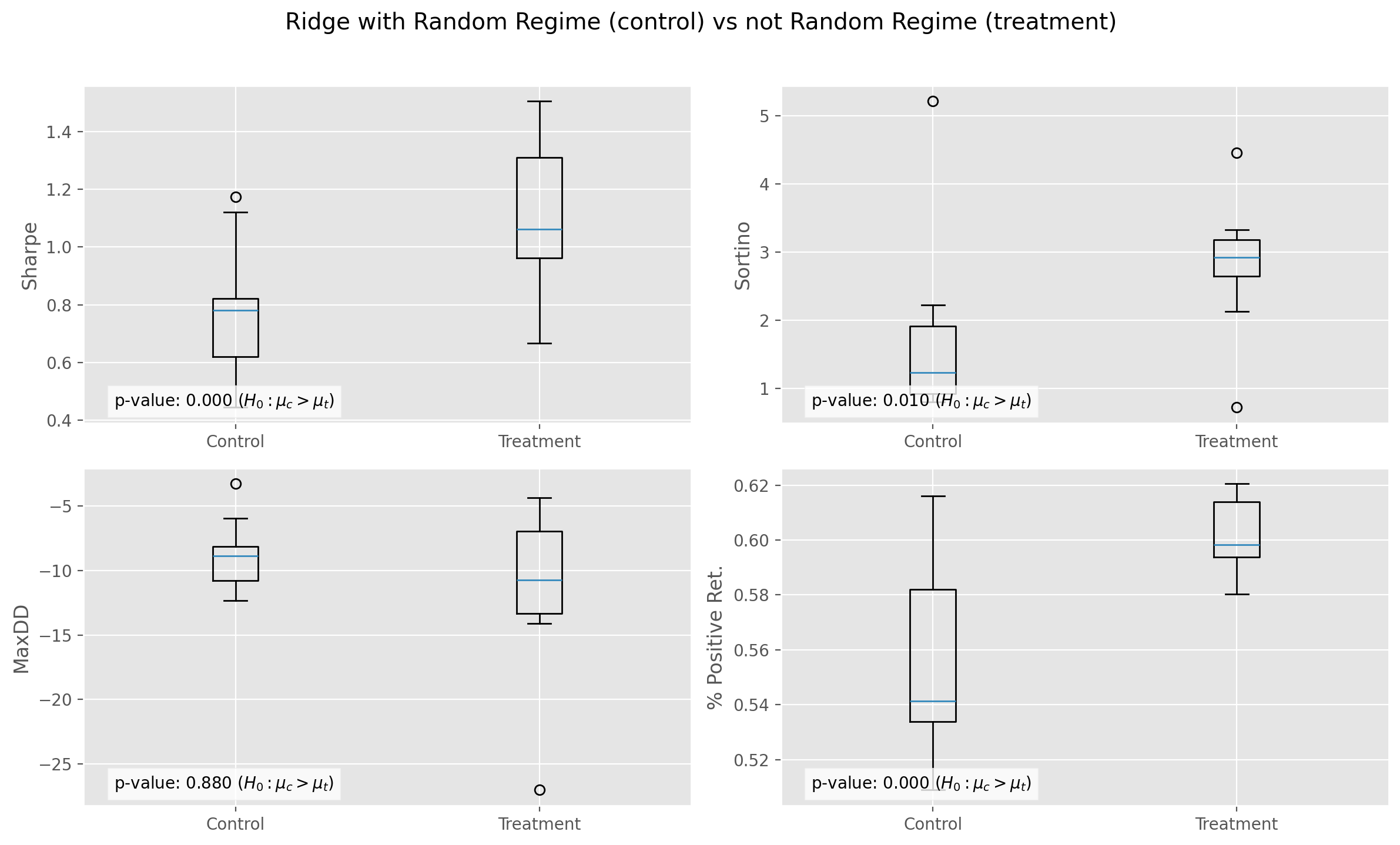}
    \caption{The figure depicts boxplots for controls and treatment portfolio metrics. The control is defined as the ridge (Ridge) method based on random regimes, whereas the control is defined as the ridge method based on not random regimes. The figure also shows p-values for the t-test for two related samples where the null hypothesis is that the control mean is higher than the treatment mean.}
    \label{fig:fig9}
\end{figure}

Figures \ref{fig:fig7}, \ref{fig:fig8}, and \ref{fig:fig9} present comparative boxplots of performance metrics for control (random regimes) versus treatment (non-random regimes) across our three modeling approaches. Each figure displays four key metrics: Sharpe ratio, Sortino ratio, maximum drawdown (MaxDD), and percentage of positive returns. The boxplots are accompanied by p-values from t-tests for two related samples where the null hypothesis states that the control mean is higher than the treatment mean. Table \ref{tab:table3} complements these visualizations by providing statistical evidence through the Nemenyi test, quantifying the significance of differences between random and non-random regime approaches using rank statistics.

The naive forecasting method with non-random regimes (Figure \ref{fig:fig7}) generally performs better and is more stable on the appropriate metrics, particularly for Sharpe and Sortino ratios. This visual pattern is confirmed by Panel A in Table \ref{tab:table3}, which shows statistically significant improvements in Sortino ratio (p-value = 0.019) and percentage of positive returns (p-value = 0.046), with a marginal improvement in Sharpe ratio (p-value = 0.082).

Figure \ref{fig:fig8} compares MVO (control) and Black-Litterman with regime-based views (treatment). The boxplots reveal less pronounced differences than the naive approach, except for the percentage of positive returns. This observation aligns with Panel B of Table \ref{tab:table3}, where the most significant improvement appears in the percentage of positive returns (p-value = 0.003). At the same time, other metrics show more modest or insignificant differences.

The linear ridge regression results (Figure \ref{fig:fig9}) demonstrate the most striking improvements when using non-random regimes. The treatment group shows notably higher and more concentrated Sharpe ratios, Sortino, and percentage of positive returns supported by Panel C of Table \ref{tab:table3}. Statistical results present substantial evidence to reject both null hypotheses for both metrics. Regarding the t-test, we estimate p-values of 0.004, 0.01, and 0.013 for the Sharpe, Sortino, and percentage of positive returns, respectively. This means that we can strongly reject the null hypothesis that the average metric for the ridge model that uses random regimes is higher than the one that uses not random regimes. Similar evidence is found using the Nemenyi test, where we rejected the null that there is no difference between the two models in terms of Sharpe, Sortino, and percentage of positive returns.

These findings suggest structured regime information consistently enhances the quality of returns across all three approaches, with the most substantial benefits observed in the linear ridge regression framework. The improvement in Sharpe ratios and percentage of positive returns indicates that regime information helps models better capture market opportunities.

Second, the pattern of improvements suggests that regime information primarily enhances return generation capabilities rather than downside risk management, as evidenced by the generally insignificant changes in maximum drawdown across all approaches.

Third, the varying degrees of improvement across methods indicate that more sophisticated models, particularly the linear ridge regression, may be better equipped to leverage regime information effectively. This suggests that the choice of modeling framework plays a crucial role in maximizing the benefits of regime-based information in portfolio management.

\begin{table}[H]
\centering
\caption{The table shows results for the Nemenyi test. Control and treatment methods are defined as in Figure \ref{fig:fig7}-\ref{fig:fig9}. The table shows the rank statistics for control and treatment methods, as long as p-values are associated with it.}
\begin{tabular}{lllll}
\hline
Metric & Control Rank & Treatment Rank & p-value \\
\hline
\multicolumn{4}{l}{\textit{Panel A: Naive Random vs not Random Regimes Portfolios}} \\
Sharpe & 1.333 & 1.667 & 0.082* \\
Sortino & 1.167 & 1.833 & 0.019** \\
MaxDD & 1.500 & 1.500 & 0.480 \\
\% Positive Ret. & 1.250 & 1.750 & 0.046** \\
\hline
\multicolumn{4}{l}{\textit{Panel B: BL Random vs not Random Regimes Portfolios}} \\
Sharpe & 1.333 & 1.667 & 0.068* \\
Sortino & 1.500 & 1.500 & 0.295 \\
MaxDD & 1.667 & 1.333 & 0.394 \\
\% Positive Ret. & 1.083 & 1.917 & 0.003*** \\
\hline
\multicolumn{4}{l}{\textit{Panel C: Ridge Random vs not Random Regimes Portfolios}} \\
Sharpe & 1.000 & 2.000 & 0.000*** \\
Sortino & 1.167 & 1.833 & 0.010*** \\
MaxDD & 1.583 & 1.417 & 0.880 \\
\% Positive Ret. & 1.083 & 1.917 & 0.000*** \\
\hline
\multicolumn{4}{l}{\textit{Note:} ***, **, * indicate significance at 1\%, 5\%, and 10\% levels}
\end{tabular}
\label{tab:table3}
\end{table}

\begin{figure}[H]
    \centering
    \includegraphics[width=0.75\textwidth]{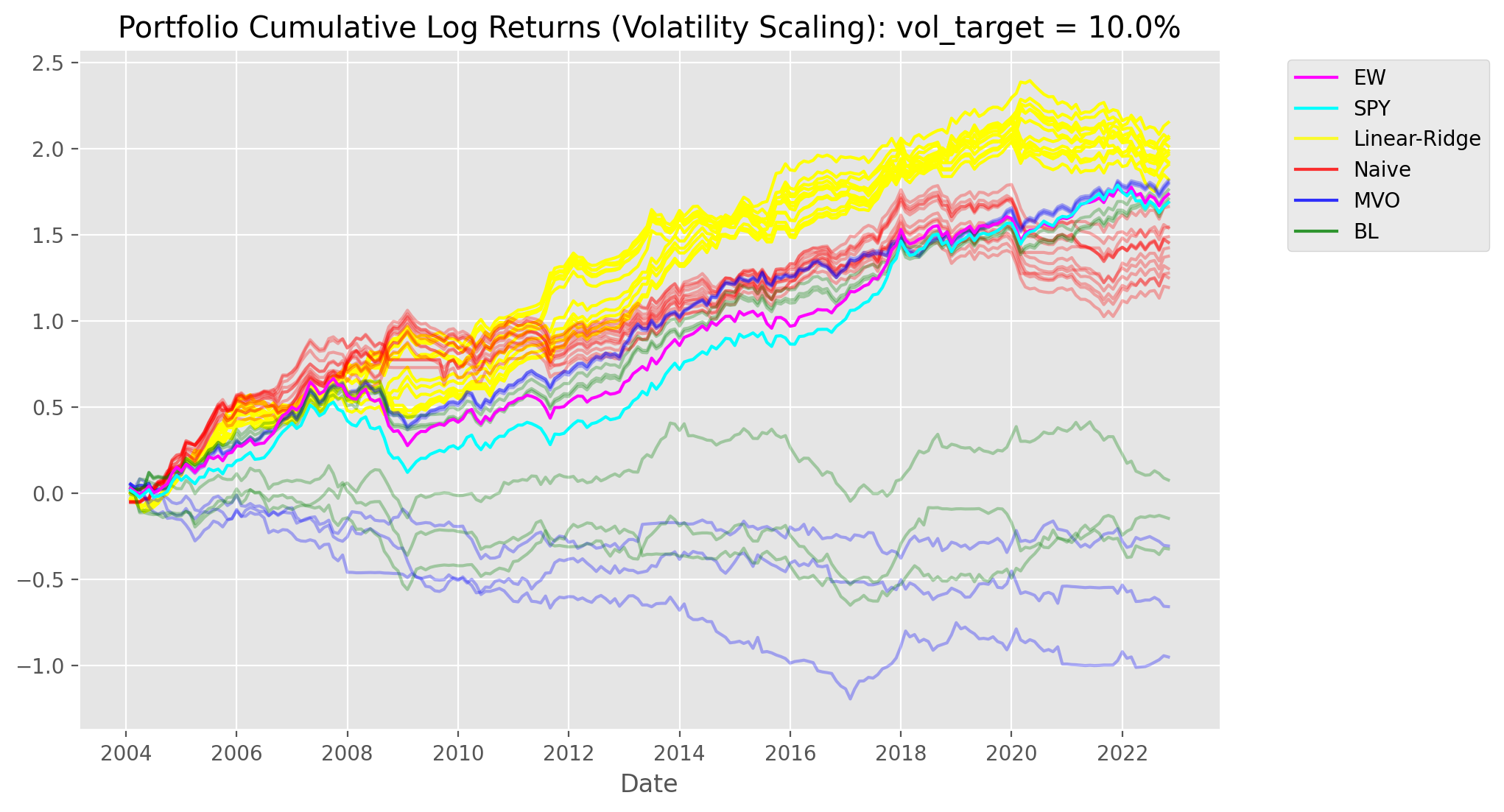}
    \caption{Cumulative log returns with 10\% volatility scaling of various tactical allocation methods over time. Linear ridge strategies are in yellow, naive strategies are in red, mean-variance optimization strategies are in blue, and Black-Litterman strategies are in green.}
    \label{fig:fig10}
\end{figure}

\begin{figure}[H]
    \centering
    \includegraphics[width=0.75\textwidth]{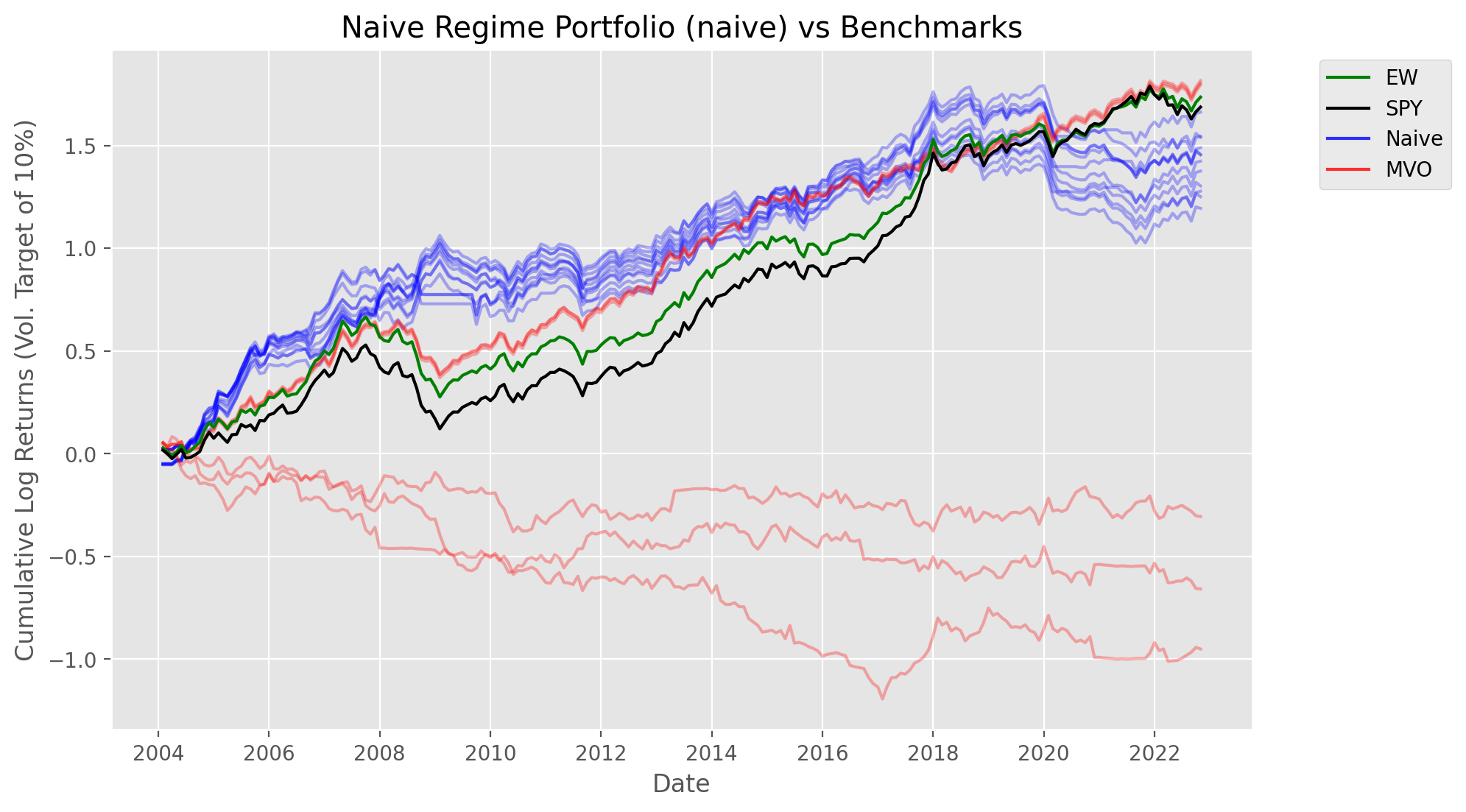}
    \caption{Cumulative log returns with 10\% volatility scaling of various tactical allocation methods over time. Naive strategies are in blue, and mean-variance optimization strategies are in red.}
    \label{fig:fig11}
\end{figure}

\begin{figure}[H]
    \centering
    \includegraphics[width=0.75\textwidth]{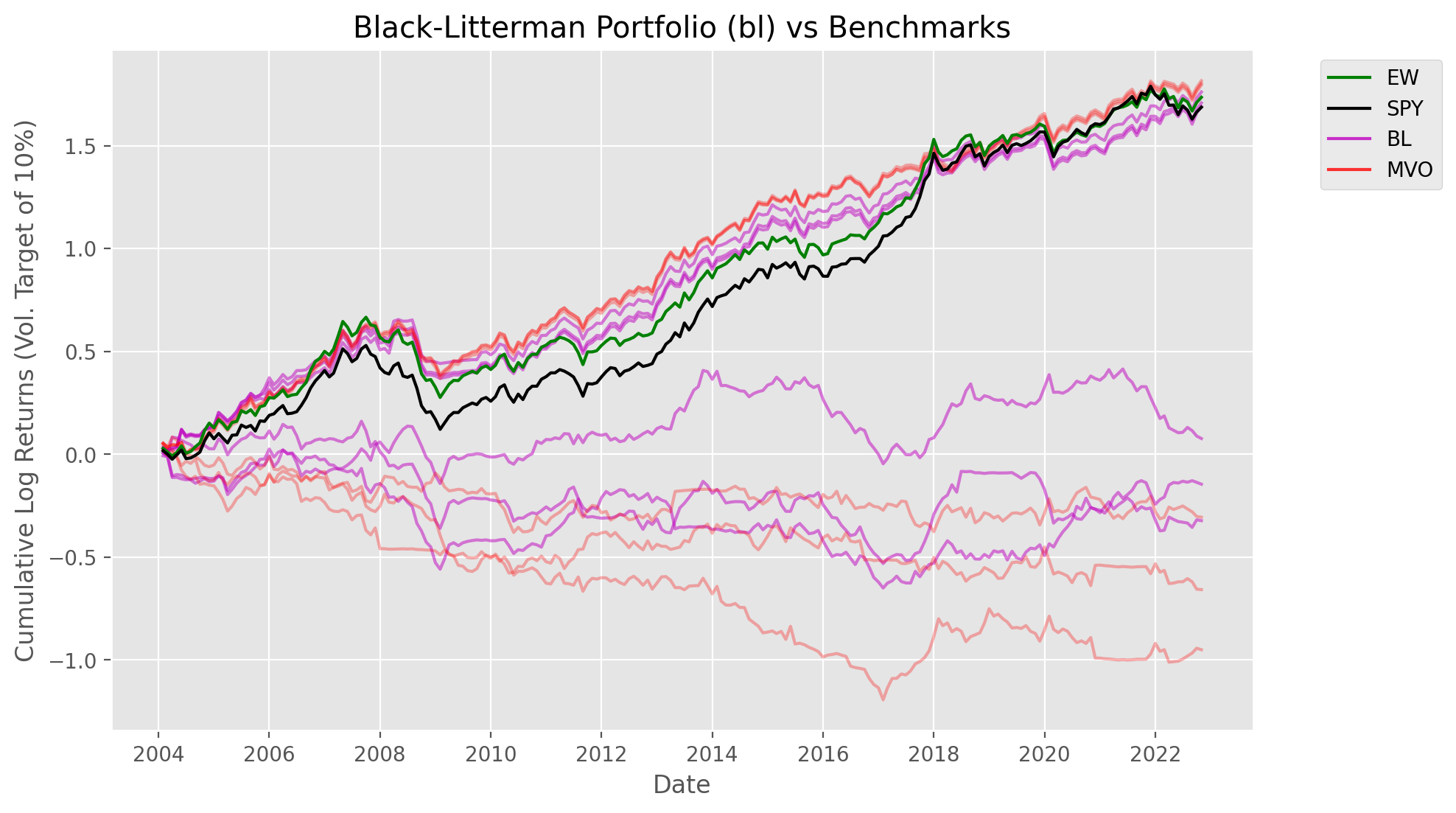}
    \caption{Cumulative log returns with 10\% volatility scaling of various tactical allocation methods over time. Black-Litterman strategies are in pink, and mean-variance optimization strategies are in red.}
    \label{fig:fig12}
\end{figure}

\begin{figure}[H]
    \centering
    \includegraphics[width=0.75\textwidth]{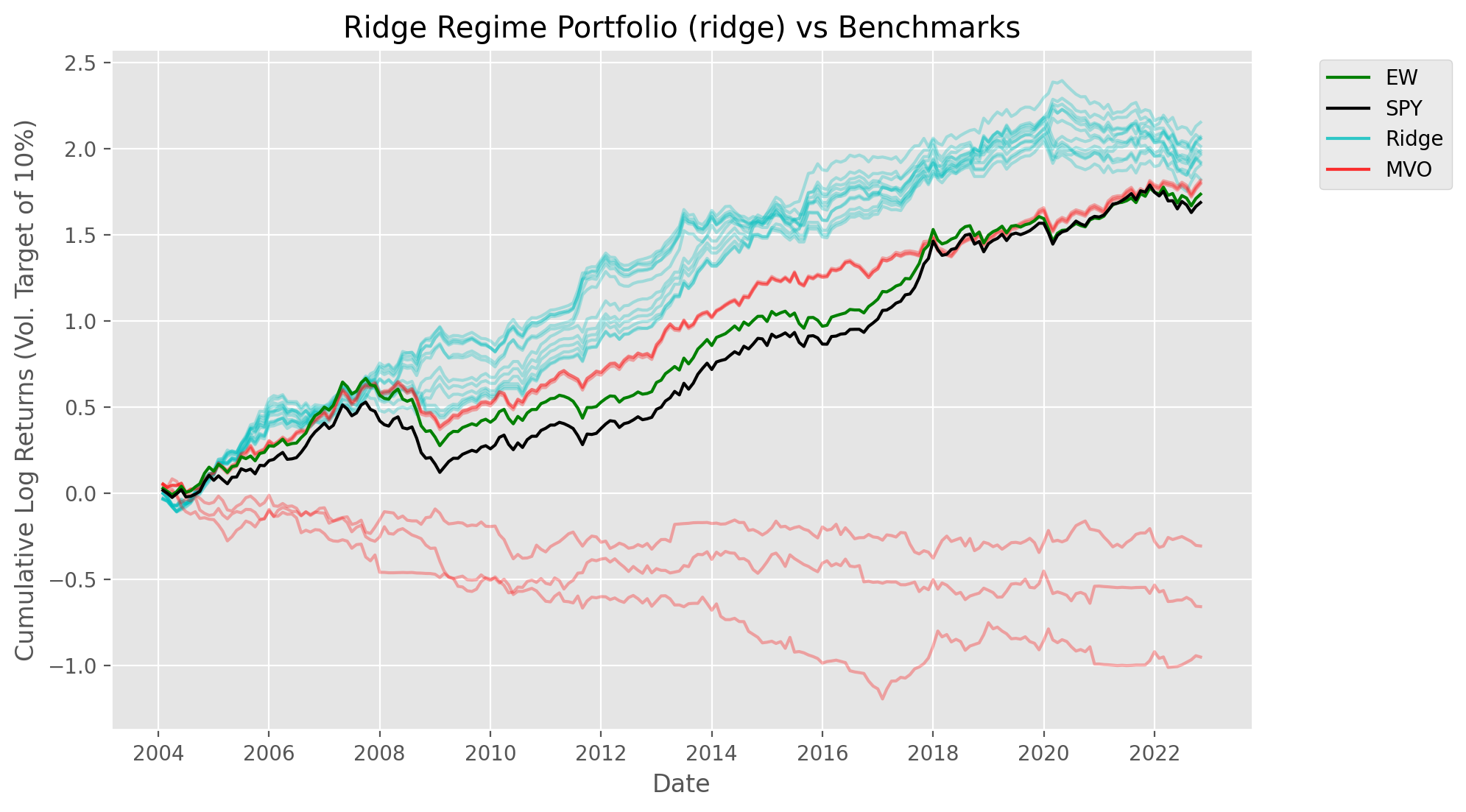}
    \caption{Cumulative log returns with 10\% volatility scaling of various tactical allocation methods over time. Linear ridge strategies are in blue, and mean-variance optimization strategies are in red.}
    \label{fig:fig13}
\end{figure}

Examining Figures \ref{fig:fig10}-\ref{fig:fig13} and Tables \ref{tab:table4}-\ref{tab:table6}, we observe several notable performance patterns. Among naive strategies, naive\_lo\_4 achieves a Sharpe ratio of 1.065 and exhibits the lowest maximum drawdown (-6.719) within its category. The naive portfolios consistently outperform the SPY benchmark during major market downturns, as shown in Figure \ref{fig:fig11}, though long-short variants demonstrate higher volatility.

In the Black-Litterman implementations, bl\_lo\_2 records the highest Sharpe ratio (1.177) and superior positive return percentage (0.665) compared to MVO strategies (0.629). However, MVO strategies achieve better Sortino ratios, with mvo\_lo\_4 reaching 2.132. Figure \ref{fig:fig12} shows BL strategies tracking closer to market benchmarks than MVO approaches, while long-short variants of both methods frequently produce negative Sharpe ratios (e.g., mvo\_lns\_2 at -0.516).

Again, the linear ridge regression approach produces the strongest overall metrics. The ridge\_lo\_3 configuration achieves the highest Sharpe ratio (1.505) across all strategies, while ridge\_lo\_2 demonstrates exceptional downside protection with a Sortino ratio of 4.449 and the lowest maximum drawdown (-4.389) among all approaches.

Our results underscore both the challenge and potential of active portfolio management. The benchmark indices (SPY and EW) demonstrate remarkably consistent performance with 66.2\% positive returns, highlighting the difficulty of outperforming passive strategies. This robust benchmark performance makes the success of certain regime-based strategies particularly noteworthy.

The linear ridge regression approach, especially in its long-only implementations, stands out by achieving superior risk-adjusted returns. With a Sharpe ratio of 1.505 and exceptional downside protection, these results suggest that machine-learning approaches may be particularly well-suited to capturing regime dynamics. This outperformance is especially meaningful given the high hurdle set by market benchmarks.

A striking pattern across all strategies is the superior performance of long-only implementations compared to their long-short counterparts. This finding challenges conventional wisdom about the diversification benefits of short positions and suggests that the added complexity may introduce more noise than signal in regime-based strategies. Similarly, lower-dimensional models (l=2 or l=3) consistently outperform higher-dimensional ones (l=4), indicating that simpler specifications may better capture fundamental market dynamics.

While the Black-Litterman approach does not achieve the highest absolute performance, it demonstrates an interesting balance between active management and benchmark tracking. Its tendency to align closely with market benchmarks while still achieving competitive Sharpe ratios suggests potential value for investors seeking enhanced passive strategies.

These findings carry significant implications for practical implementation. While outperforming market benchmarks remains a formidable challenge, our results suggest that thoughtfully implemented regime-based strategies, particularly those utilizing machine learning approaches, can achieve this goal. The key element is to not increase model complexity but make informed choices about implementation, favoring simpler, long-only approaches that focus on capturing fundamental market dynamics. While not universal across all configurations, this success in outperforming challenging benchmarks demonstrates the potential value of regime-based approaches in portfolio management.

{\footnotesize
\begin{table}[H]
  \centering
  \caption{Performance metrics for portfolio strategies. Bold values indicate the best performance per column.}
  \begin{tabular}{lllll}
      \toprule
      Model & Sharpe & Sortino & MaxDD & \% Positive Ret. \\
      \midrule
      naive\_lns\_2 & 0.651 & 1.984 & -14.942 & 0.580 \\
      naive\_mx\_2 & 0.773 & 1.771 & -14.643 & 0.585 \\
      naive\_los\_2 & 0.796 & 1.742 & -14.595 & 0.580 \\
      naive\_lo\_2 & 0.981 & 2.532 & -6.163 & 0.540 \\
      naive\_lns\_3 & 0.402 & \textbf{3.533} & -16.148 & 0.558 \\
      naive\_mx\_3 & 0.740 & 1.948 & -14.681 & 0.576 \\
      naive\_los\_3 & 0.568 & 2.357 & -14.773 & 0.558 \\
      naive\_lo\_3 & 0.899 & 2.630 & \textbf{-5.859} & 0.536 \\
      naive\_mx\_4 & 0.746 & 1.486 & -17.807 & 0.589 \\
      naive\_lns\_4 & 0.734 & 1.207 & -17.699 & 0.580 \\
      naive\_los\_4 & 0.788 & 1.618 & -17.064 & 0.580 \\
      naive\_lo\_4 & 1.065 & 2.541 & -6.719 & 0.549 \\
      \midrule[\heavyrulewidth]
      mvo\_lns\_2 & -0.516 & -0.601 & -35.806 & 0.455 \\
      mvo\_lo\_2 & 1.128 & 2.022 & -6.776 & 0.629 \\
      mvo\_lns\_3 & 0.058 & 0.102 & -12.390 & 0.513 \\
      mvo\_lo\_3 & 1.104 & 1.963 & -6.846 & 0.629 \\
      mvo\_lns\_4 & -0.510 & -0.723 & -43.622 & 0.446 \\
      mvo\_lo\_4 & \textbf{1.129} & 2.132 & -6.933 & 0.629 \\
      spy & 0.818 & 1.331 & -33.492 & \textbf{0.662} \\
      ew & 0.838 & 1.445 & -32.286 & \textbf{0.662} \\
      \bottomrule
  \end{tabular}
  \label{tab:table4}
\end{table}
}

{\footnotesize
\begin{table}[H]
  \centering
  \caption{Performance metrics for portfolio strategies with Black-Litterman and MVO approaches. Bold values indicate the best performance per column.}
  \begin{tabular}{lllll}
      \toprule
      Model & Sharpe & Sortino & MaxDD & \% Positive Ret. \\
      \midrule
      bl\_lns\_2 & 0.390 & 1.507 & -17.926 & 0.522 \\
      bl\_lo\_2 & \textbf{1.177} & 1.744 & -8.636 & 0.665 \\
      bl\_lns\_3 & 0.023 & 0.023 & -26.081 & 0.518 \\
      bl\_lo\_3 & 1.071 & 1.306 & -10.612 & \textbf{0.670} \\
      bl\_lns\_4 & 0.096 & 0.130 & -24.185 & 0.487 \\
      bl\_lo\_4 & 1.107 & 1.479 & -9.620 & 0.656 \\
      \midrule[\heavyrulewidth]
      mvo\_lns\_2 & -0.516 & -0.601 & -35.806 & 0.455 \\
      mvo\_lo\_2 & 1.128 & 2.022 & \textbf{-6.776} & 0.629 \\
      mvo\_lns\_3 & 0.058 & 0.102 & -12.390 & 0.513 \\
      mvo\_lo\_3 & 1.104 & 1.963 & -6.846 & 0.629 \\
      mvo\_lns\_4 & -0.510 & -0.723 & -43.622 & 0.446 \\
      mvo\_lo\_4 & 1.129 & \textbf{2.132} & -6.933 & 0.629 \\
      spy & 0.818 & 1.331 & -33.492 & 0.662 \\
      ew & 0.838 & 1.325 & -32.286 & 0.662 \\
      \bottomrule
  \end{tabular}
  \label{tab:table5}
\end{table}
}

{\footnotesize
\begin{table}[H]
  \centering
  \caption{Performance metrics for linear ridge regression and MVO approaches. Bold values indicate the best performance per column.}
  \begin{tabular}{lllll}
      \toprule
      Model & Sharpe & Sortino & MaxDD & \% Positive Ret. \\
      \midrule
      ridge\_lns\_2 & 0.666 & 0.718 & -27.020 & 0.621 \\
      ridge\_mx\_2 & 1.015 & 2.865 & -12.300 & 0.612 \\
      ridge\_los\_2 & 0.876 & 2.127 & -9.926 & 0.621 \\
      ridge\_lo\_2 & 1.134 & \textbf{4.449} & -4.389 & 0.607 \\
      ridge\_lns\_3 & 0.953 & 2.977 & -13.266 & 0.594 \\
      ridge\_mx\_3 & 1.330 & 2.674 & -7.795 & 0.598 \\
      ridge\_los\_3 & 1.051 & 3.322 & -10.976 & 0.621 \\
      ridge\_lo\_3 & \textbf{1.505} & 3.170 & -4.386 & 0.594 \\
      ridge\_lns\_4 & 0.966 & 3.067 & -14.090 & 0.580 \\
      ridge\_mx\_4 & 1.303 & 2.554 & -10.470 & 0.589 \\
      ridge\_los\_4 & 1.074 & 2.813 & -13.546 & 0.598 \\
      ridge\_lo\_4 & 1.490 & 3.201 & \textbf{-4.368} & 0.594 \\
      \midrule[\heavyrulewidth]
      mvo\_lns\_2 & -0.516 & -0.601 & -35.806 & 0.455 \\
      mvo\_lo\_2 & 1.128 & 2.022 & -6.776 & 0.629 \\
      mvo\_lns\_3 & 0.058 & 0.102 & -12.390 & 0.513 \\
      mvo\_lo\_3 & 1.104 & 1.963 & -6.846 & 0.629 \\
      mvo\_lns\_4 & -0.510 & -0.723 & -43.622 & 0.446 \\
      mvo\_lo\_4 & 1.129 & 2.132 & -6.933 & 0.629 \\
      spy & 0.818 & 1.331 & -33.492 & \textbf{0.662} \\
      ew & 0.838 & 1.325 & -32.286 & \textbf{0.662} \\
      \bottomrule
  \end{tabular}
  \label{tab:table6}
\end{table}
}

\section{Conclusion and Future Work}
By leveraging clustering techniques from machine learning, we have developed a novel data-driven approach to regime detection using macroeconomic data. We show how it uncovers insights related to market dynamics and use our regime detection mechanism to derive a regime transition graph that tracks well with existing economic intuitions. Lastly, we show how these regimes can be used to augment forecasting models by applying a probabilistic condition on historical data.

More research on this topic could explore how other more sophisticated clustering techniques could be used for financial regime detection. The question of how the choice of macroeconomic features affects the quality of the resulting cluster classifications remains. Finally, further work in this direction could also examine how these regimes could be used to improve other financial analytic processes.

\begin{acks}
We thank the Oxford-Man Institute of Quantitative Finance (OMI) for supporting this work and providing the resources necessary to make this research possible.
\end{acks}

\bibliographystyle{ACM-Reference-Format}
\bibliography{sample-base}
\end{document}